\documentclass[useAMS,usenatbib]{mn2e}
\usepackage{graphicx}
\usepackage{amssymb}
\usepackage[english]{babel}

\usepackage[dvipsnames]{color}

\usepackage[dvipsnames]{color}

\def\ltsima{$\; \buildrel < \over \sim \;$}
\def\simlt{\lower.5ex\hbox{\ltsima}}
\def\gtsima{$\; \buildrel > \over \sim \;$}
\def\simgt{\lower.5ex\hbox{\gtsima}}

\newcommand{\kmps}{km s\ensuremath{^{-1} }}
\newcommand{\Msun}{M\ensuremath{_\odot}}
\newcommand{\Oo}{\displaystyle}

\title[To be or not to be oblate.]{To be or not to be oblate: the shape of the dark matter halo in the polar ring galaxies.}

\author[Khoperskov et al.]{
S.A. Khoperskov$^{1,2}$\thanks{khoperskov@inasan.ru},  A.V. Moiseev$^{3,2}$,  A.V. Khoperskov$^4$, A.S. Saburova$^2$\\
$^1$Institute of Astronomy, Russian Academy of Sciences\thanks{The system of Russian Academy of Sciences institutes was liquidated on Sep 2013}, Pyatnitskaya st., 48, 119017 Moscow, Russia\\
$^2$Sternberg Astronomical Institute, Moscow M.V. Lomonosov State University, Universitetskij pr., 13, 119992 Moscow, Russia\\
  $^3$Special Astrophysical Observatory, Russian Academy of Sciences,  369167 Nizhnii Arkhyz,  Karachaevo-Cherkesskaya Republic,  Russia\\
 $^4$Volgograd State University, Universitetsky pr., 100, 400062 Volgograd, Russia}

\date{}

\pagerange{\pageref{firstpage}--\pageref{lastpage}} \pubyear{2014}

\def\LaTeX{L\kern-.36em\raise.3ex\hbox{a}\kern-.15em
    T\kern-.1667em\lower.7ex\hbox{E}\kern-.125emX}

\begin{document}

\label{firstpage}

\maketitle

\begin{abstract}
With the aim to determine the spatial
distribution of the dark matter halo, we investigate two polar ring galaxies  NGC 4262 and SPRC-7. For both galaxies  the stellar kinematics data for the central galaxy were obtained from optical spectroscopy at the 6-m telescope of the Special Astrophysical Observatory of the Russian Academy of Sciences. The information about polar gaseous components  was taken from the  optical 3D-spectroscopic observations of  ionized gas (for SPRC-7) and H~{\sc i} radio observations (for NGC 4262). SPRC-7 is the  system with a relative angle $\delta=73^{\circ}$ towards the central galaxy and a quite massive stellar-gaseous polar component. Meanwhile NGC~4262 is the classic polar case with $\delta=88^{\circ}$ where the polar ring  mainly consists of neutral gas with a negligible stellar contribution to the mass. We are hence dealing with two different systems and the results are quite diverse too.  The observed properties of both galaxies were compared with the results of self consistent simulations of velocity fields of the polar component along with the rotation curve of the central lenticular galaxy.
For SPRC-7 we have found a slightly flattened halo towards the polar plane with the axis ratio $c/a \simeq 1.7 \pm 0.2$ for the isothermal halo model and $c/a \simeq 1.5 \pm 0.2$ for the NFW model. The case of NGC~4262 is more unusual, the shape of the dark matter distribution varies strongly with  radius. Namely, the dark matter halo is fattened in the vicinity of the galactic disc ($c/a \approx 0.4 \pm 0.1$), however it is prolate far beyond the central galaxy ($c/a \approx 1.7$ for the isothermal halo and $c/a \approx 2.3$ for NFW).
\end{abstract}

\begin{keywords}
Galaxies: evolution; galaxies: haloes; galaxies: kinematics and dynamics;
\end{keywords}

\section{Introduction}
 Following the $\Lambda$CDM model, dark matter (DM) is a crucial factor of evolution on the spatial scales of galactic clusters and large-scale structure of the entire Universe. 
The role of DM in the dynamics and evolution of individual  galaxies is much less clear.
However, the presence of dark matter haloes around the galaxies is obvious. The long-term stability of galactic discs is supported by the presence of a massive collisionless spheroidal component~\citep{1981AJ.....86.1791B,1991MNRAS.249..523B,2013A&A...557A.131M}. The low ratio of the radial velocity dispersion $\sigma_r$ to the maximum disc rotation velocity $V_{\max}$ also indicates the existence of the massive halo~\citep*{2003ARep...47..357K}.
The evolution  and lifetime of galactic bars strongly depends on the interaction with the dark matter halo~\citep{2006ApJ...637..567S,2006ApJ...639..868S,2013arXiv1312.1690A,2013MNRAS.429.1949A}. Dark matter influences the dynamics of tidal streams in our Galaxy~\citep{2001ApJ...551..294I,2004ApJ...610L..97H,2014MNRAS.437..116B} and dwarf satellites. The thickness of the stellar disc   closely  depends on the  disc-to-halo mass ratio~\citep*{1991PAZh...17..884Z,2010AN....331..731K,2013MNRAS.434.2373R}, moreover, the observed vertical equilibrium  is explained by the presence of a massive DM halo. The investigation of  the gaseous and stellar kinematics allows to estimate the total  amount of  DM mass and its spatial distribution inside the galaxies and their neighbourhood. The measured kinematics of the  the Milky Way thick disc stars allows to conclude that the volume density of the dark matter in the Solar vicinity is insignificant~\citep*{2011MNRAS.416.2318G}. 

Triaxiality of dark matter haloes  is confirmed owing to the observations of various galactic systems: anisotropy of weak-lensing signals~\citep{2012A&A...545A..71V,2013AIPC.1534..156S}, dynamics of hypervelocity of stars~\citep{2005ApJ...634..344G}, flaring of the gas layer~\citep{2000MNRAS.311..361O}, gravitational lensing~\citep{2000ApJ...538L.113N} and so on.  The shapes of the X-ray  isophotes  for the some elliptic galaxies pointed out the manifestation of a triaxial dark matter halo~\citep{2002ApJ...577..183B}. Dynamical features of  the Sagittarius Stellar Tidal Stream in our Galaxy might be explained by the triaxial dark matter distribution~\citep*{2009ApJ...703L..67L,2013MNRAS.434.2971D}. It is believed that  galactic warps  are usually formed and  maintained within the triaxial DM haloes~\citep{1995ApJ...442..492D,2006MNRAS.370....2S}. It was also shown that the nonstationarity of the galactic spiral pattern rotation might be the result of interactions with a non-axisymmetric dark matter halo~\citep{2013MNRAS.431.1230K,2013arXiv1310.7021V}. From the theoretical point of view, the triaxial halo shape  is supported, by  numerous  cosmological simulations which  demonstrate the formation of   DM halos with different  spatial scales $a\ne b\ne c$ at almost any condition~\citep{1991ApJ...378..496D,2002ApJ...574..538J,2005ApJ...627..647B,2006MNRAS.367.1781A, 2007ApJ...671.1135K,2010MNRAS.407..435A}.

\begin{figure}
\includegraphics[width=1\hsize]{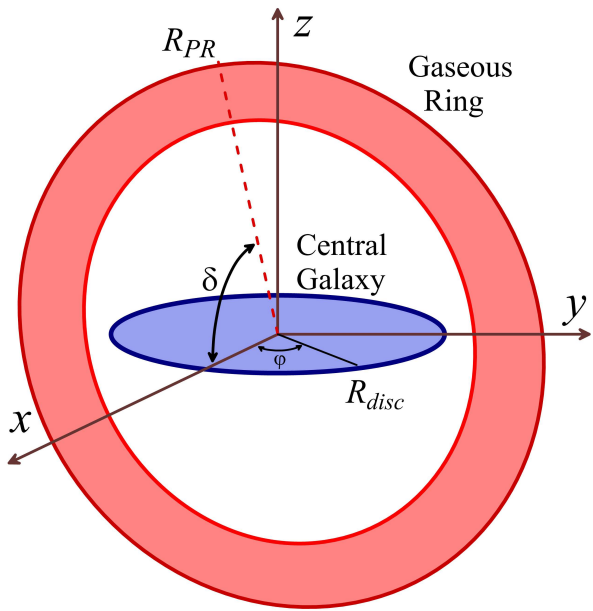}
\caption{Scheme of spatial orientation of the components in a polar ring galaxy. We placed the central lenticular galaxy for both PRGs in the XY-plane. A gaseous ring is situated in the YZ-plane in the classic polar case for NGC~4262 --- $\delta = 88^{\circ}$. The ring plane in SPRC-7 is inclined  towards the XY-plane at the angle $\delta = 73^{\circ}$. }\label{fig1::scheme}
\end{figure}

 Galaxies having  polar rings are among the most ambitious candidates for the  measurements of the DM halo shape. This is a type of peculiar objects, characterized by two kinematically decoupled components: a host (or central) galaxy and a polar component (disc or ring), rotating in different planes (see the Fig.~\ref{fig1::scheme}). The first catalogue, containing 157  PRG candidates and related objects was  created by~\citet{1990AJ....100.1489W}.
Recently, more than two hundred galaxies with polar rings were found in the SDSS~\citep{2011MNRAS.418..244M}, some of which have a kinematic confirmation. It was believed that polar structures are the result of a close interaction of galaxies~\citep{1990AJ....100.1489W,2012MNRAS.425.1967S},
accretion of companion's matter~\citep{1997A&A...325..933R} or infalls of cold gaseous filaments~\citep{2009ApJ...696L...6S,2010ApJ...714.1081S}.
In any case, this kind of multi-spin galaxies is a good  tool  to study the dark and baryonic
matter gravitational potential around them. Central galaxies of PRGs are mainly represented by the S0 or elliptical galaxies, which indicates the absence of a significant amount of gas inside of them.
Statistical investigation of PRGs revealed a small amount ($\approx 6 \%$) of inclined systems~\citep{2013arXiv1311.4138S}. This fact proves that polar (or closely resembling) structures should be stabilized by the gravitational potential of the halo.

It should be noted that the internal structure of PRGs is poorly studied. There are some proofs of ongoing star formation within the gaseous ring~\citep*{1994A&A...290..693R}. Non-axisymmetric perturbations from the central galaxies~(similar to bar-like perturbations in the discs of galaxies) and self gravity for massive polar rings should lead to the formation of spiral structures~\citep*{2006A&A...446..905T}.  The large-scale non-axisymmetric halo potential is an effective generator of  spiral structures in the gaseous component~\citep{2002ApJ...574L..21B,2012ARep...56...16K}. However, there is no satisfactory observational evidence of these substructures in PRGs.


The most well-studied PRG is NGC~4650A.
During numerous studies the understanding of DM halo shape in this galaxy has significantly evolved. The first halo shape estimates were done  via the comparison of maximal velocities of the central galaxy~(hereinafter denoted as CG) and the polar component~\citep{1987ApJ...314..439W}. This method allows to establish the oblate DM halo shape towards the galactic plane with the halo semi-axis relation $c/a \sim 0.83 \pm 0.21$.

The position of  polar ring galaxies on the Tully-Fisher diagram allows us to estimate the deviations of the gravitational potential from the spherical shape. The majority of PRGs do not follow the main relation for the spiral galaxies in the Tully-Fisher diagram  because they are shifted to larger rotation velocities~\citep*{2003ApJ...585..730I,2004A&A...416..889R,2013A&A...554A..11C}. A comparison with the model galaxies reveals an oblate DM halo towards the polar plane~\citep{2008ASPC..396..483I,2010AIPC.1240..379I}. 

A kinematic model based on the variations of the $M/L$ ratio and halo flattening provides the constrains on the dark halo shape in NGC~4650A~\citep{1990ApJ...361..408S}.   If the ring is not massive,  the DM halo is non-spherical, however, the halo shape varies in the range of E0-E6 (halo flattened towards the polar plane). Another kinematic model by \citet{1996A&A...305..763C} also  argues that the dark halo of the polar ring galaxy NGC~4650A is flattened towards the polar ring plane. Nevertheless, despite the differences in the estimated shape of the DM halo, an important argument in favour of the flattened halo in the PRG (in the plane of the galaxy) is the stability of  polar orbits for such a kind of configuration of the gravitational potential~\citep{1982ApJ...263L..51S,1993AJ....105.1745W}.

A surprising feature of DM distribution was { found in} the simulations of the formation of PRGs due to major merging. \citet{2012MNRAS.425.1967S} showed that the major and minor halo axes are swapped along the radius. In fact, this means a change of the type of the halo shape. In the  central part the halo could be flattened to the plane of the central galaxy, but beyond the optical radius of the galaxy it becomes flattened to the polar plane.  Thus, within this framework  it is possible to combine the accumulated knowledge about the NGC 4650A   halo parameters. Nevertheless, there are no observations in favour of such kind of dark matter distribution   yet. 

Formally, there are no observational (or theoretical) evidences confirming that the PRG assembly strongly differs from   regular galaxies. If it is true, the DM halo structure is expected to be  similar. However, the resulting baryonic objects are not similar. Of course, dark matter plays  the dominant role in the formation of galaxies. This means that the current galaxy type depends on the abundance of gas (and/or stars) inside the merging DM haloes at $z>0$. Perhaps it is possible to form various galaxies within the given DM haloes assembly process. On another hand, the DM halo shape in the cosmological context is not constant in time tending to the more spherical shape at smaller redshifts \citep[e.g. see][]{2006MNRAS.367.1781A}. As it was mentioned above,  polar rings around the galaxies are usually more stable in non-spherical haloes. Thus, the proper evolution of DM haloes should destroy the polar components. Thus, PRGs haloes have another structure, than the DM around regular galaxies because it is related to another epoch of   evolution. However, the mutual influence of the DM halo and the baryonic fraction could complicate these submission. We are not sure now that the conclusions about the DM halo shape of PRGs could be directly applied to the regular galaxies.

In this paper we present the observations of two PRGs together with the study of the spatial distribution of dark matter. The following observations were used: the long-slit spectral data of the central galaxies NGC 4262 and SPRC-7  obtained at the 6-m telescope of the Special Astrophysical Observatory of the Russian Academy of Sciences (SAO RAS); 21 cm H~{\sc i} observations of the ring in NGC 4262~\citep{2010MNRAS.409..500O}; the scanning Fabry-Perot interferometric H$\beta$ observations of ionized gas in the ring of SPRC-7 \citep{2011MNRAS.418.1834F}. The halo shape determination is based on the detailed decomposition of the baryon matter rotation in different planes, namely, the stellar discs in central galaxies and the gaseous rings. Our paper  is organized as follows:  Section 2  describes the observations of both galaxies. Section 3  presents the model of the galactic potential, basic parameters of the system and the fitting procedure. Basic results are presented in Section 4.

\section{Observational Data}

\subsection{Long-slit spectral observations}

The spectroscopic observations at the prime focus of the SAO RAS 6-m telescope  were  made with the SCORPIO multi-mode focal reducer~\citep{AfanasievMoiseev2005} and  its new version SCORPIO-2 \citep{2011BaltA..20..363A}. When operated in the long-slit mode, both devices have the same slit,  6.1 arcmin in height,  with a scale of 0.36 arcsec per pixel. However, with a similar spectral resolution SCORPIO-2 provides a twice larger spectral range. The CCDs employed were the EEV 42-40 in the SCORPIO and E2V 42-90  in the SCORPIO-2.

\begin{figure*}
\centerline{
\includegraphics[height=0.5\textwidth]{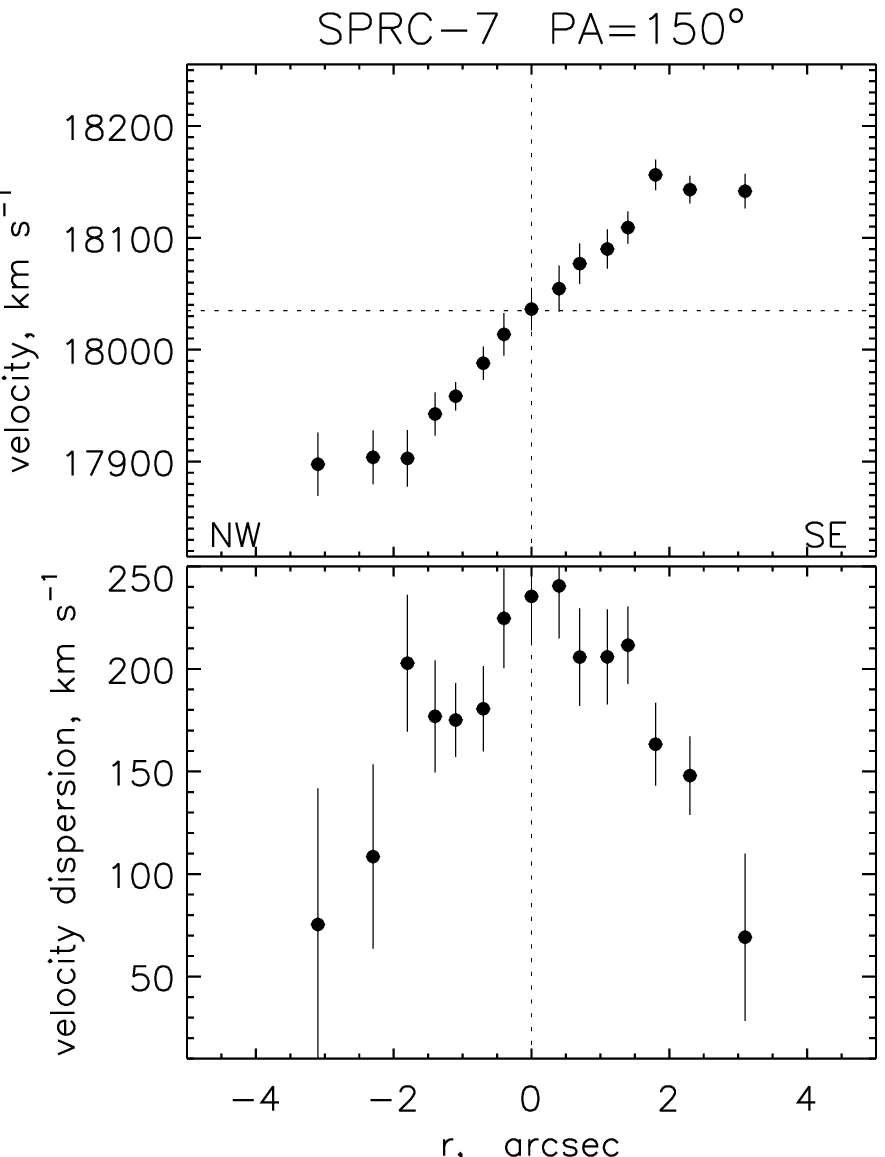}
\includegraphics[height=0.5\textwidth]{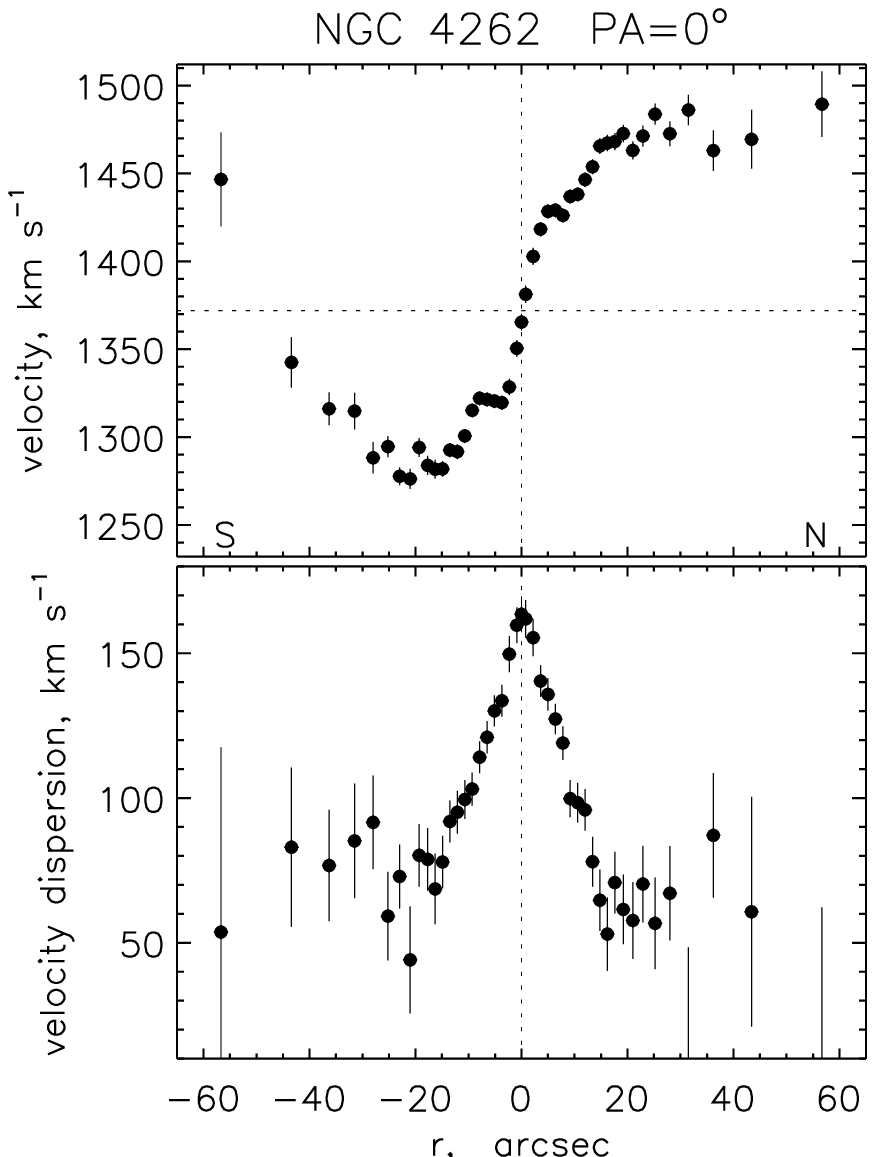}
\includegraphics[height=0.5\textwidth]{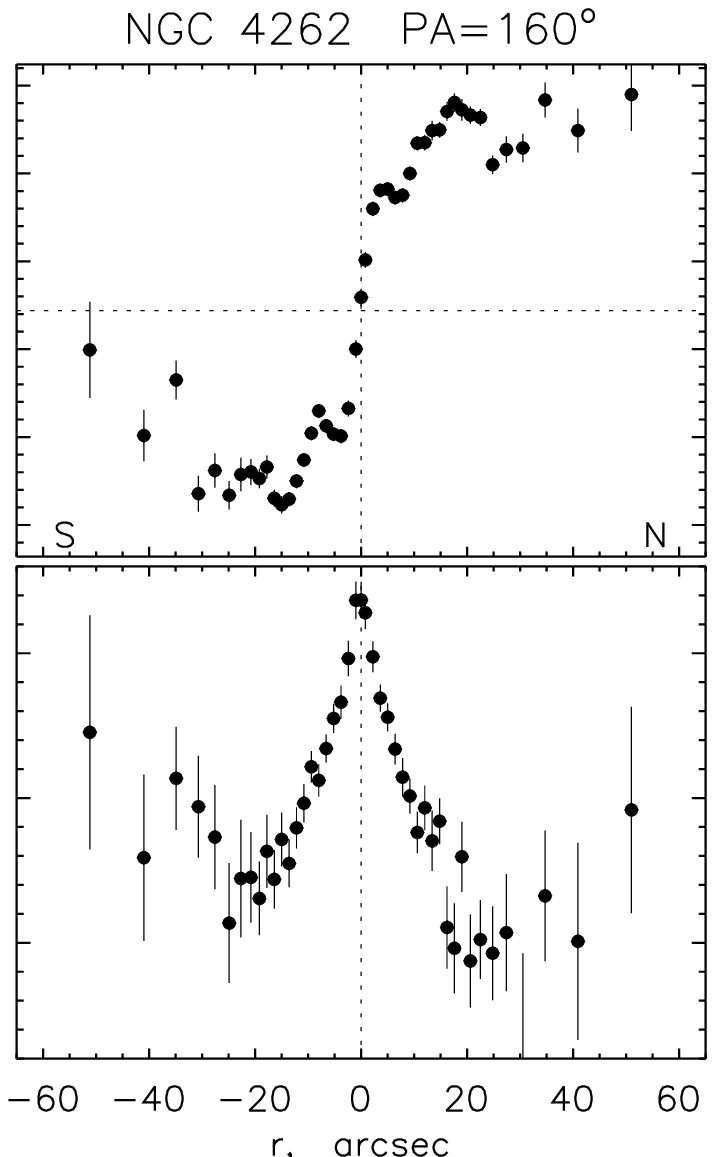}
}
 \caption{The results of the SAO RAS 6-m  telescope long-slit observations: line-of-sight velocities and velocity dispersion of stars along the major axis. The dotted lines mark the position of nucleus and accepted systemic velocity. 
 } \label{fig_longslit}
\end{figure*}

Table 1 gives the log of observations:   position angles of the spectrograph slit for each galaxy, observing date, slit width, spectral range, spectral resolution (estimated by the mean FWHM of airglow lines), total exposure, and seeing.  The data reduction  was made  in a standard way using the IDL-based software package developed at the SAO RAS   \citep{AfanasievMoiseev2005}. The measurements of the distribution of the line-of-sight velocities  and stellar velocity dispersion  were carried out by cross-correlating the spectra of galaxies with the spectra of the template star observed on the same nights. The measurement technique has already been described in our previous papers \citep{Moiseev2001,2011MNRAS.418..244M}. Figure~\ref{fig_longslit} shows the line-of-sight velocity and velocity dispersion distributions for each slit position.

\begin{table*}
\caption{Log of the observations}\label{tab_obs}
\begin{tabular}{lrrccccc}
\hline
Galaxy    & Slit $PA$ & Date                 & Slit width & Sp. range         &  Sp. resol. & Exp. time    &  Seeing  \\
             &   (deg)&                                &   (arcsec)  &   (\AA)         &   (\AA)   &  (min)          &   (arcsec)\\
\hline
  SPRC-7       & 150        &  5 Apr 2011   &   1.0         & 3900--5700 &    5.0     & $120$         &   1.5\\ 
NGC~4262   &  0           & 3 Apr 2013     &   0.7         & 3700--7240 &    3.7     & $140$         &  3.1 \\ 
                    &  160        & 6 Apr 2013     &   0.5          & 3700--7240 &    3.0     & $100$         &  3.0 \\ 
   \hline
\end{tabular}
\end{table*}

\begin{figure*}
\centerline{
\includegraphics[width=0.5\textwidth]{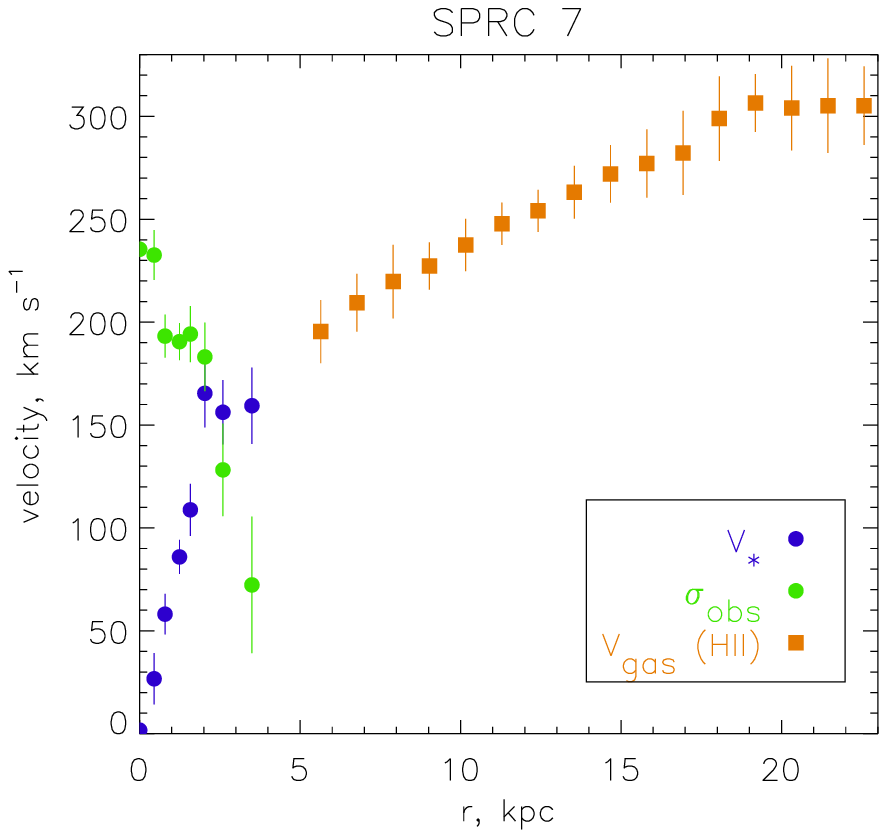}
\includegraphics[width=0.5\textwidth]{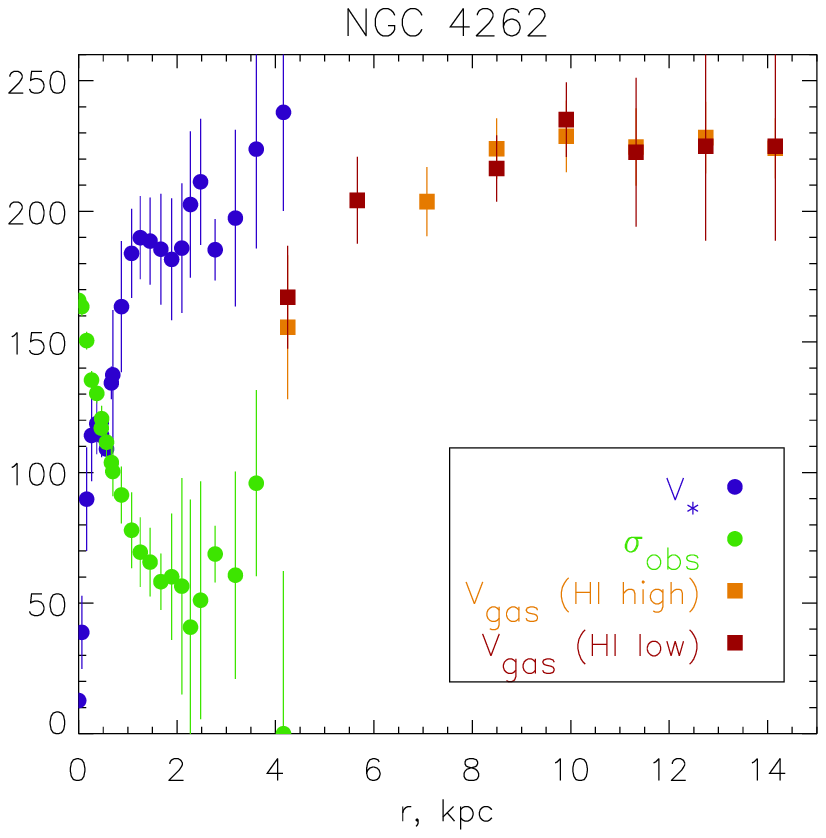}
}
\caption{The observed rotation curves of the stars and gas components and mean line-of-sight velocity dispersion radial distribution in  SPRC~7 (left) and NGC 4262 (right). The `high' and `low' data points correspond to the H~{\sc i} data  with  $66\times17$  and $79\times30$ arcsec beam size. }
\label{fig_obs_RC}
\end{figure*}


\begin{figure*}
\centerline{
\includegraphics[width=0.5\textwidth]{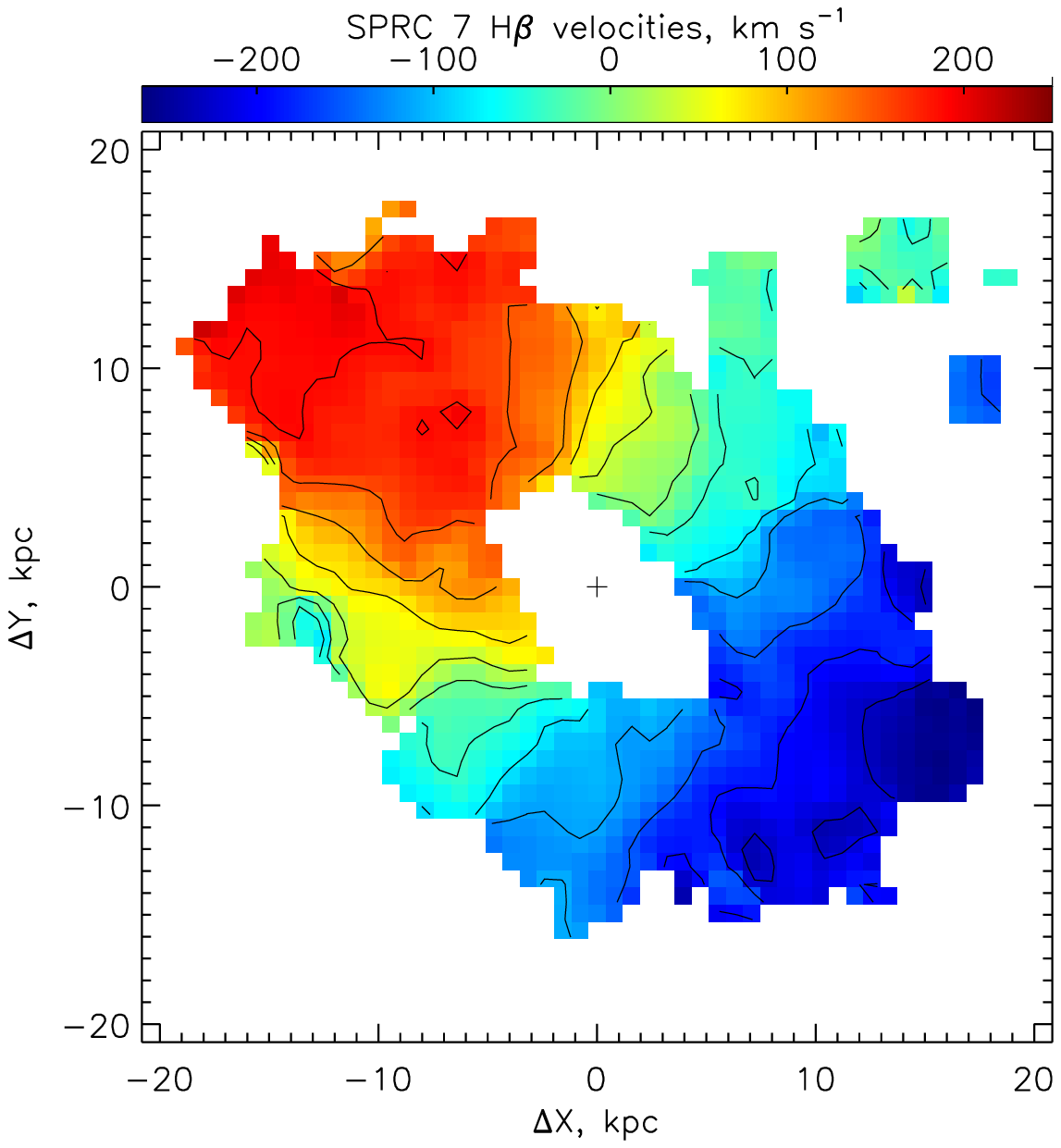}
\includegraphics[width=0.5\textwidth]{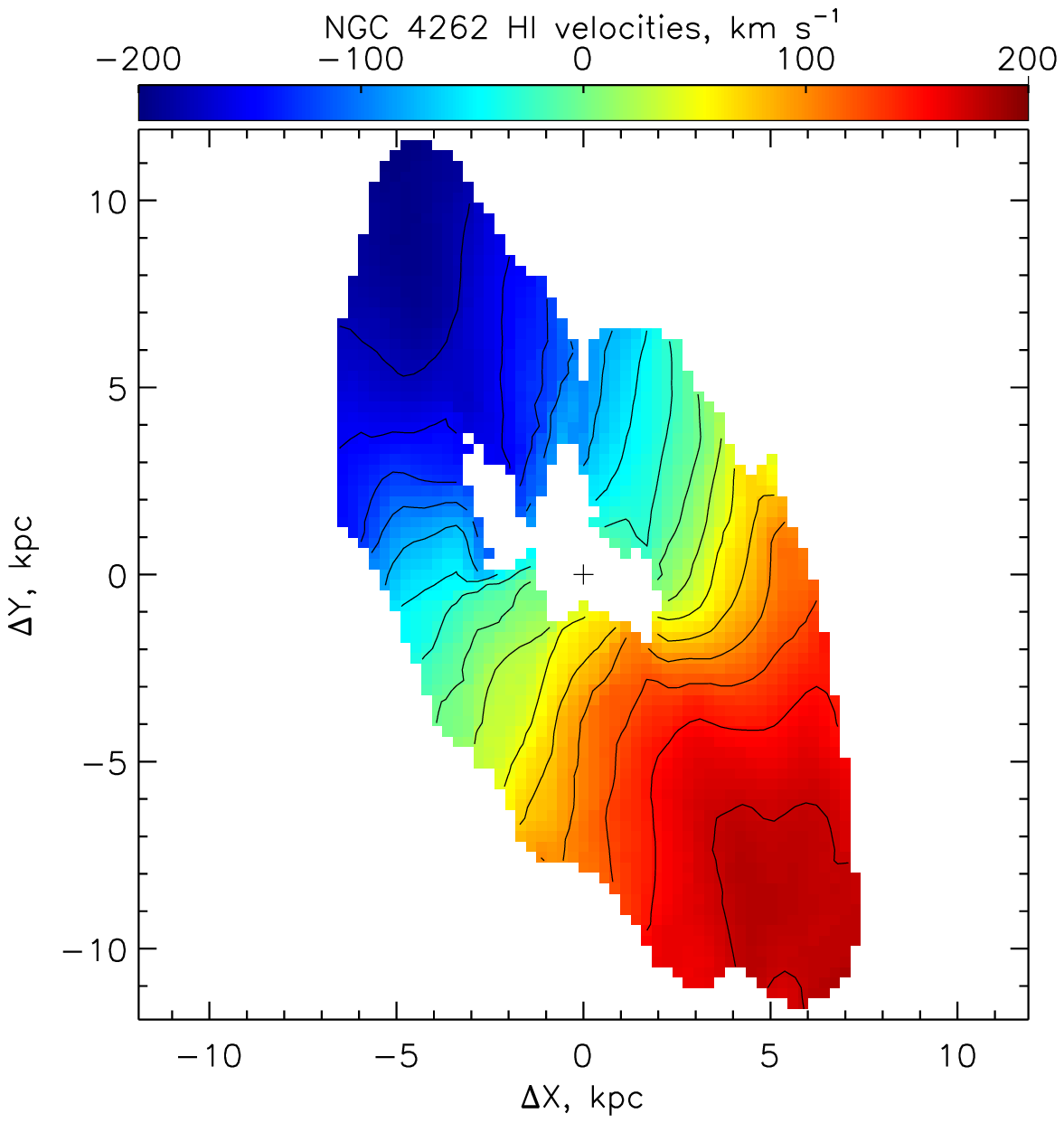}
}
 \caption{The line-of-sight velocity fields  of the polar components $(V)$, after subtracting the systemic velocities: for the ionized gas in SPRC 7 (left) and for the H~{\sc i} in NGC 4262 (right). The cross marks the rotation centre position. North is upwards, east  is to the left.}
\label{fig_velfields}
\end{figure*}

\begin{figure*}
\centerline{
\includegraphics[width=0.5\textwidth]{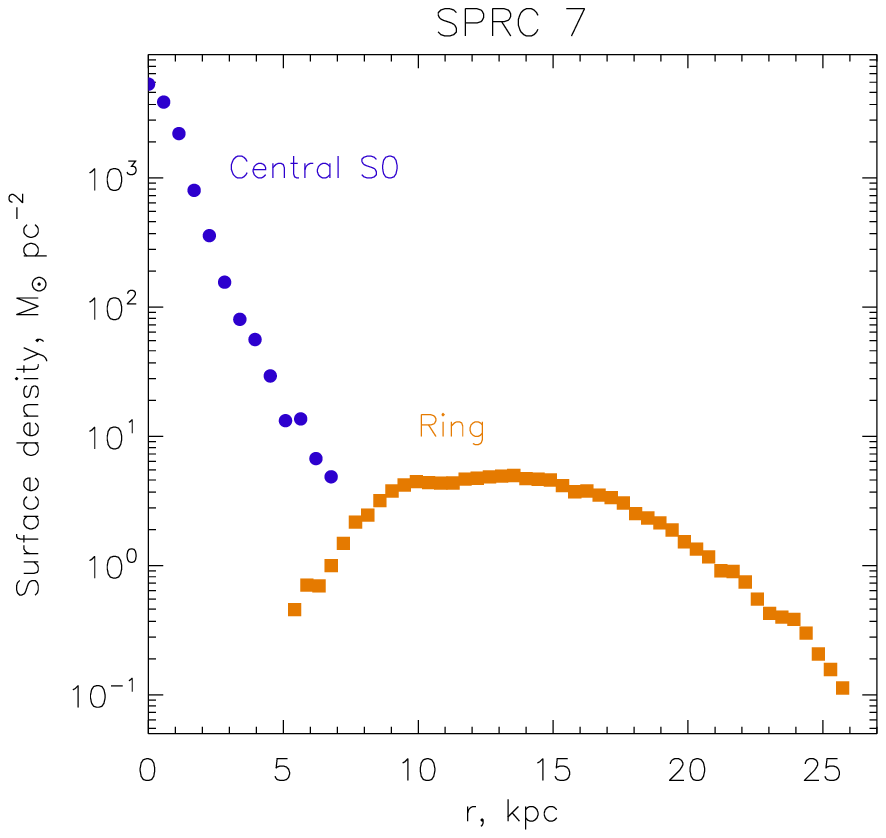}
\includegraphics[width=0.5\textwidth]{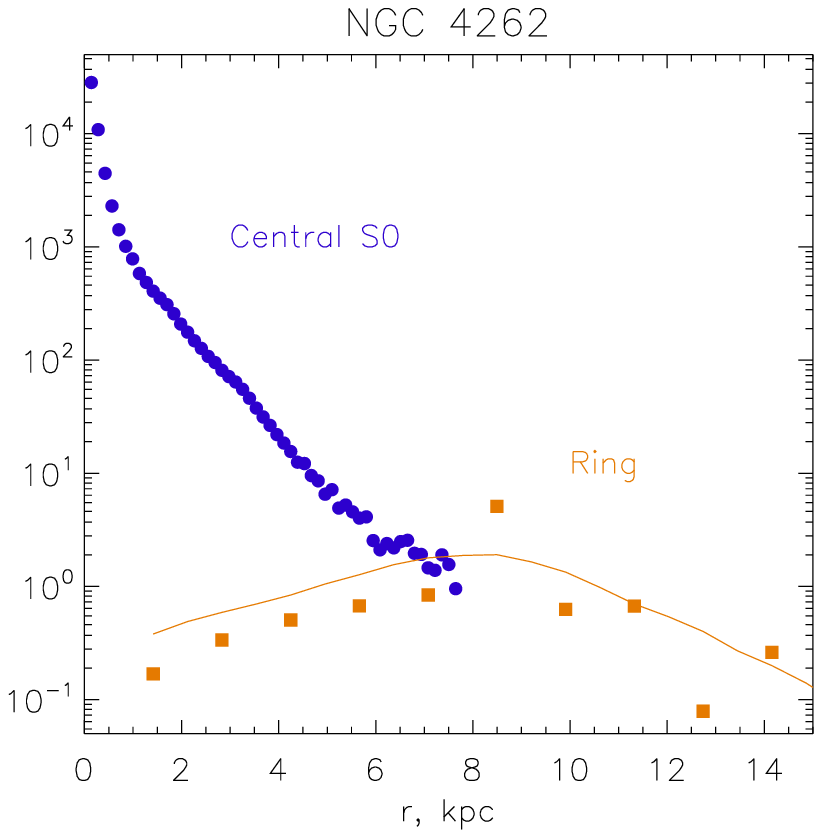}
}
\caption{The azimuthally averaged mass density distribution for the internal and external components. In the case of SPRC 7 (left)    both curves correspond to the  mass density of stars derived  from the SDSS data (see the text). In the NGC 4262 (right) the stellar mass was estimated only for the central galaxy. For the ring the figure shows the surface density of the gas ($1.4\times\mbox{M}_{\mbox{H{\sc i}}}$) derived from the TiRiFiC model (orange symbols) and an azimuthally averaged density map (the orange curve). }
\label{fig_galmass}
\end{figure*}

\subsection{SPRC 7}

The polar structure in this galaxy has been discovered and described  in the paper by \citet{Brosch2010} based on the SDSS images and 6-m telescope spectroscopic data.  On the optical images the galaxy appears as a giant counterpart of NGC~4650A because the central early-type galaxy is surrounded by a large ring with a significant contribution of young stars. In contrast to NGC~4650A, the   SPRC~7 galaxy is more distant, huge (about  40 kpc in diameter) and  moderately inclined. The external structure is more likely to be a `polar disc' than a ring, the total luminosity in the \textit{g}-band is even larger than  for the central galaxy. The accepted distance for the galaxy is $D=233$~Mpc, the scale is $1.13$~kpc per arcsec.

Our  long-slit spectra were taken along the photometric major axis of the CG. The new measurements of the stellar kinematics parameters show a significantly smaller  scatter  compared  with data presented in \citet{Brosch2010}. The line-of-sight velocities, as well as stellar velocity dispersion have a symmetric  distribution relative to the galaxy centre  (Figure~\ref{fig_longslit}). The averaged stellar rotation curve and velocity dispersion distribution are shown in Figure~\ref{fig_obs_RC}. The   following orientation parameters were accepted: the line-of-nodes position angle $PA_{CG}=150\pm5$ deg, the inclination $i_{CG}=49\pm10$ deg. These values,   based on the analysis of  the  SDSS $r$-band image slightly differ from the values calculated in~\citet{Brosch2010}  from the stellar kinematics  ($127\pm4$ and $50\pm15$~deg, respectively). We think that the morphological $PA_{CG}$ is more reliable in comparison with the kinematic one due to the large scatter of stellar velocities presented along both slit positions in \citep{Brosch2010}.

For the external ring we have accepted the   $PA_{ring}=49\pm3$ deg, and the inclination of $i_{ring}=50\pm10\degr$ according to \citet{Brosch2010} estimation from the circular model fitting of the ionized-gas velocity field. The corresponding H$\beta$ velocity field is shown in  (Figure~\ref{fig_velfields}), the inner part was masked because of a strong contamination from the stellar absorption \citep[see Fig.~7 in][]{Brosch2010}, the rotation curve is presented in   Figure~\ref{fig_obs_RC}.

The present combination of $(PA_{ring}, i_{ring})$ and $(PA_{CG}, i_{CG})$ provides two solutions \citep[eq. (1) in][]{2008AstBu..63..201M} for the mutual angle between the central galaxy and polar ring (denoted as $\delta$ on the Fig.~\ref{fig1::scheme}): $73\pm12\degr$ and $58\pm9\degr$. We accepted the first value as a more stable case (i.e. {the closest} to the polar plane).

The surface density distributions of the stellar mass were calculated separately for each component using the $g$ and $r$ SDSS DR8 images averaged in the ellipses with the corresponding $PA$ and $i$ (see Fig.~\ref{fig_galmass}). We use the foreground Galactic extinction  values from the NED\footnote{http://ned.ipac.caltech.edu} and  K-corrections from \citet{2010MNRAS.405.1409C}. The mass-to-luminosity ratios $(M/L)$ were calculated in each radii using $(g-r)$ color according  to \cite{2003ApJS..149..289B}. The CG has a negative gradient of $r$-band $M/L$ from $\sim7$ in the nucleus to $2$ at the periphery. While in the blue ring this value is constant of the order of $0.5$ $M_\odot/L_\odot$.

\subsection{NGC 4262}

NGC 4262 is a well-known lenticular barred galaxy in the Virgo cluster, the accepted distance is 14.6 Mpc according to the \textit{HST} photometry \citep{2005ApJ...634.1002J} that corresponds to the scale of 0.071 kpc per arcsec. In contrast to the previous case, the polar component in NGC 4262  has no  significant stellar mass contribution, while the galaxy is surrounded by an extended gaseous ring discovered in the 21 cm radio data  by \citet*{1985A&A...144..202K}.  \citet*{Bettoni2010} have shown that: (i)  the UV \textit{GALEX} images reveal the presence of a young stellar population in the ring of neutral hydrogen; (ii)  both the ionized gas in the inner region and the neutral hydrogen at large distances from the centre rotate in the plane, which is strongly inclined to the stellar disc of the galaxy.  The contamination of UV knots  in the optical SDSS images is very faint. Hence,  in our simulations we have accepted a pure gaseous disc   for this galaxy. \citet{Bettoni2010}  prefer to speak of an inclined ring, but a formal calculation of the mutual inclination angle yields two solutions: $39$ and 90 deg. The latter corresponds to the polar ring, as noted in their subsequent  work~\citep*{2011Ap&SS.335..231B}.

We have observed  the stellar kinematics of NGC 4262 in two slit positions, corresponding to two possible orientations of the CG line-of-nodes. The first one is $PA=160\degr$, what corresponds to the prolonged `plateau' in the radial profiles of ellipticity and $PA$ of the SDSS   isophotes \citep[see Figure~4 in ][]{Bettoni2010}.  However,  external isophotes turn to the value of $PA=0\degr$ (i.e. $180\degr$) together with a decrease of their ellipticity. The  stellar velocity field of the central region was taken with SAURON integral field spectrograph~\citep{SAURONIII}. Our analysis of this velocity field in the framework of  pure circular rotation approximation (`tilted-rings') also gives the kinematical $PA\approx 0\degr$ for the central region ($r<25''$ i.e. 1.8 kpc). However, the strong stellar bar can distort the velocity field  on the scales of $r\le1.5$ kpc from the centre.  Using the long-slit data we  took the stellar kinematics measurements up to $\sim4$  kpc from the nucleus. Two conclusions were obtained: (i)   velocity distributions along  both $PAs$ have the best agreement between each other for $PA_{CG}\approx160\degr$; (ii) the South half of the radial velocity curve shows an unusual decrease of the rotation curve in the external region (Figure~\ref{fig_longslit}). We suspect that this distortion may be related to the warping of this part of the  stellar disc, or perhaps to the contamination from the ring material, because the projected gas  velocities in this part of the ring (see the velocity field in Figure~\ref{fig_velfields})  are also redshifted relative to the systemic velocity.

These peculiar velocity points were excluded from the calculation of the final  averaged stellar rotation curve shown  in Figure~\ref{fig_obs_RC}.  The accepted orientation parameters are $PA_{CG}=160\pm10\degr$ and $i_{CG}=36\pm4\degr$. The last value was obtained from our  approximation of the isophote ellipticity  at $r=40-55$ arcsec from \citet{Bettoni2010} profiles.

For the analysis of the ring kinematics we used the results of the 21 cm WSRT radio observations presented by \citet{2010MNRAS.409..500O}. Two H~{\sc i} data cubes denoted below as `high' and 'low' with the angular resolution (beam)  of  $66\times17$  and $79\times30$  arcsec, respectively, were involved in the study. The H~{\sc i} velocity field for the high-resolution case  is shown  in the Figure~\ref{fig_velfields}. In order to take into account a strong elongation of the beam shape we  used the TiRiFiC software\footnote{http://www.astron.nl/~jozsa/tirific/},  specially designed   for the H~{\sc i} data cubes fitting \citep{TiRiFiC2007}.

Before modeling we masked the regions of the data cubes where strong noncircular motions occur. These regions are associated with active star formation visible in the  GALEX image of NGC4262. During the modeling we have assumed that the position angle and inclination are constant with radius, whereas the H~{\sc i} surface density and rotation velocity vary with galactocentric distance. We have also considered  a model, where the inclination and   position angle could vary with radius, representing a warp. However in this model we did not obtain a significant warp. The position angle and inclination were different from that found for the flat disc model only for the inner part of the disc where the evaluations were not firm due to the low density of H~{\sc i}. Furthermore, the warped model did not reproduce the data cube better than the flat disc one. Hence,   we chose the model with the constant position angle and inclination.

The obtained orientation parameters, as well as the rotation velocities  show a good agreement for the high- and low-resolution data (see the rotation curve  in  Figure~\ref{fig_obs_RC}). For the accepted $PA_{ring}=215\pm2\degr$, $i_{ring}=65\pm3\degr$ the mutual angle between CG and ring is  $\delta=50\pm6\degr$  and $\delta=88\pm6\degr$. The last value corresponds to the true polar orientation.

The radial distribution  of stellar surface density for the CG was calculated in a similar way with SPRC-7. The derived $M/L$ values show a negative gradient from 4.5 in the centre up to 1.5  $M_\odot/L_\odot$  at the periphery. For the gaseous ring  surface density distribution we used an  azimuthally averaged profile derived from the H~{\sc i}  density map. This profile has a  smoother  shape compared with the TiRiFiC results (see Figure~\ref{fig_galmass})  because  some high-density regions were excluded from the model fitting~(see above).


\begin{table*}
 \caption{Parameters of PRGs from the photometric data.}\label{Tabl2}
\begin{tabular}{lccccccccccc}
\hline
Name & $D$ & $i_{ring}$ & $i_{CG}$ & $\delta$ &   $r_d$&  $M_b$& $r_b$ &$r^{max}_{b}$ & $M_1$ & $R_{disc}$ & $R_{PR}$  \\
     & Mpc & $^{\circ} $ & $^{\circ}$ & $^{\circ}$ & 10$^{10}$~\Msun   & kpc &  10$^{10}$~\Msun & kpc & kpc &  kpc  &  kpc  \\
\hline
SPRC~7 & 261.6  & $50 \pm 10$ & $49 \pm 10$ & $58 \pm 9$ / $73 \pm 12$ & 1.77 & 0.97 &  3.4 & 0.8 & 4    & 3.7 & 23 \\
NGC~4262  & 14.6  & $65 \pm 3$ & $36 \pm 4$ & $50 \pm 6$ / $88 \pm 6$ & 1.02 & 0.9 & 0.4 & 0.2 & 1 &  4.2 & 15\\
\hline
\end{tabular}

 $D$ --- distance;
 $i_{ring}$ --- inclination angle of the central galaxy;
 $i_{CG}$ ---  inclination angle of the polar component;
 $\delta$ --- relative angle between CG and polar component;
 $r_d$ --- radial exponential scale length of the central disc;
 $M_b$ --- bulge mass;
 $r_b$ --- bulge scale;
 $r_b^{\max}$ --- bulge size;
 $M_1$ --- polar ring mass;
 $R_{disc}$ --- outer radius of the CG;
 $R_{PR}$ --- outer radius of the polar ring.

\end{table*}

\section{Modelling, decomposition and fitting}

\subsection{Modelling and decomposition}

Full set of observational data let us to use standard assumption of the asymmetric drift for simulation of central galaxy rotation curve~$V_*$, in the following form:

\begin{equation}
\Oo V_*^2 = \left(V_c^{obs}\right)^2 + \sigma^2_r \left[1 - \frac{\sigma^2_\varphi}{\sigma^2_{r}} + \frac{r}{\varrho \sigma^2_r} \frac{\partial (\varrho \sigma^2_r)}{\partial r} + \frac{r}{\sigma^2_r}\frac{\partial \langle u w \rangle}{\partial z} \right],
\end{equation}\label{eq::Jeans}
where $\Oo \frac{r}{\sigma^2_r}\frac{\partial \langle u w \rangle}{\partial z}$ is the chaotic part of the radial $u$ and vertical $w$ velocity components and this term usually equals zero. This stability criterion assumes the radial equilibrium of the disc  towards the gravitational instability.

\begin{figure*}
\includegraphics[width=0.5\hsize]{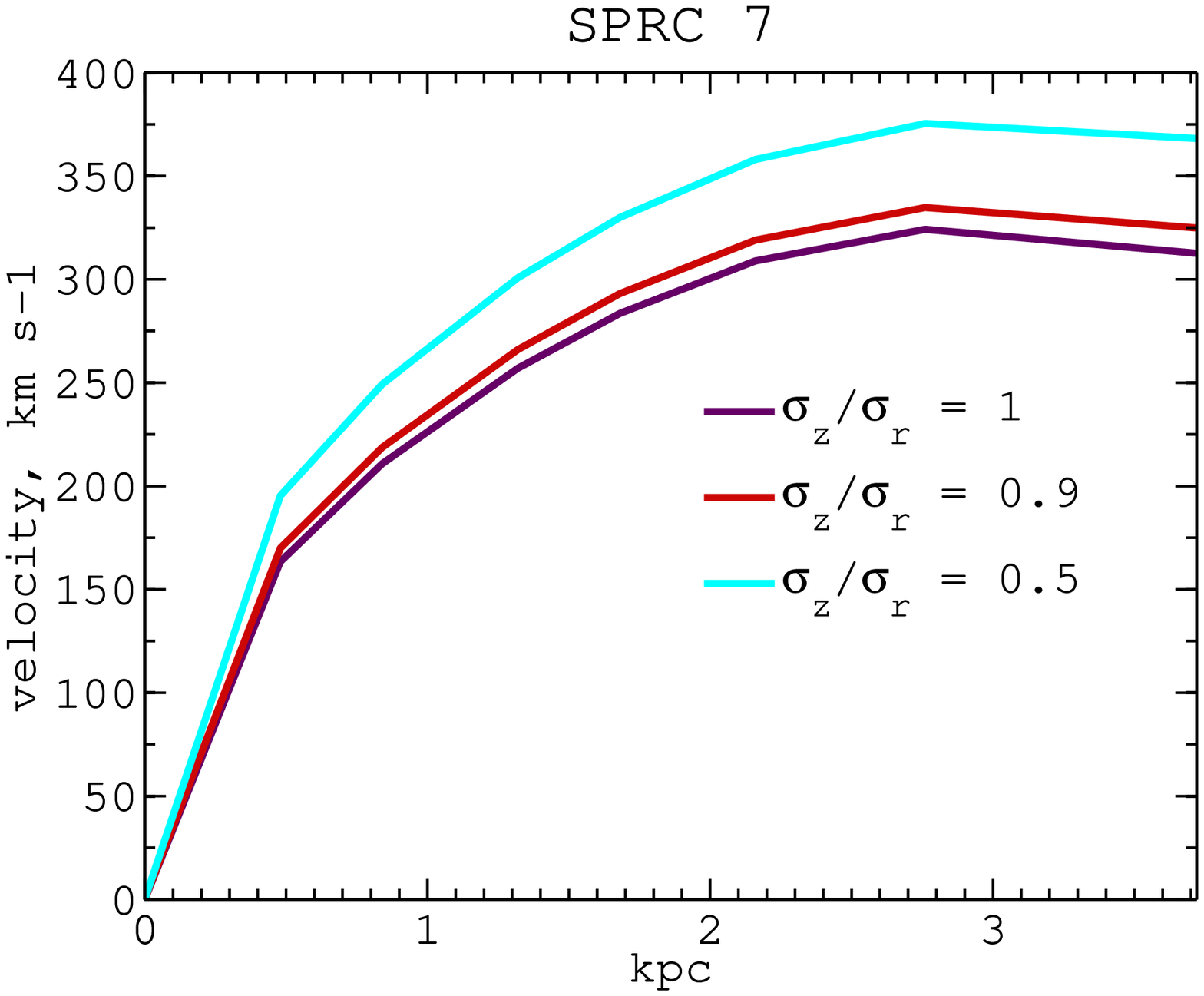}\includegraphics[width=0.5\hsize]{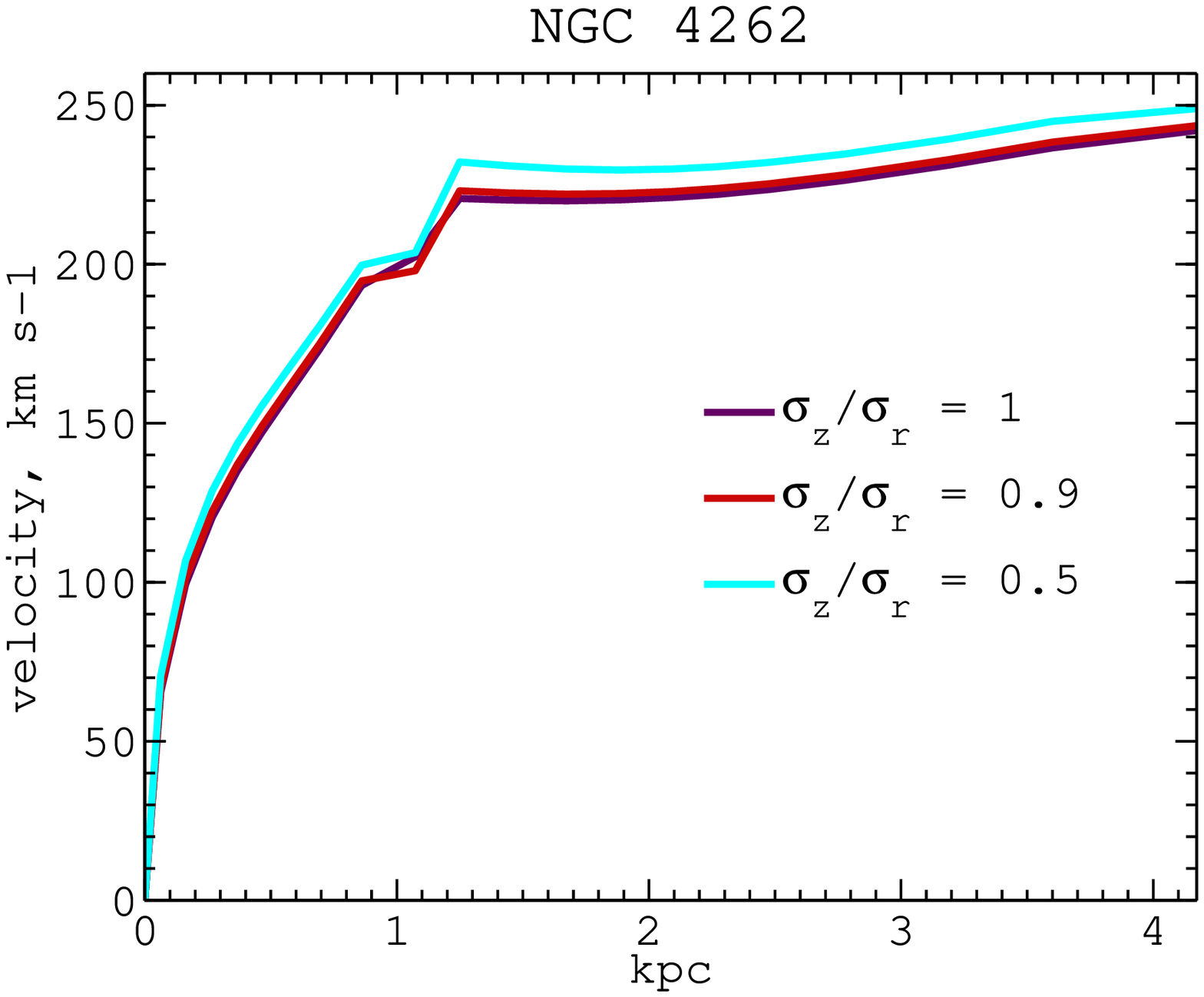}
\caption{The circular velocities $V^{obs}_c$ for SPRC~7 (left) and NGC~4262 (right) reconstructed from equation~(\ref{eq::Jeans}) with different values of $\sigma_z/\sigma_r$.}\label{fig::circular_vels}
\end{figure*}

An important issue here is the determination of velocity dispersion profiles from  the observed ones. Radial velocity dispersion  can be obtained from the measured dispersion along the major axis $\sigma_{obs}$:
\begin{equation}
\sigma_r = \sigma_{obs} \cdot \left[ (\sigma_{\varphi}/\sigma_r)^2 \sin^2 i_{CG} + (\sigma_{z}/\sigma_r)^2 \cos^2 i_{CG}  \right]\,.
\end{equation}
The first part of  the right side of this equation is determined by the Lindblad formula for the epicyclic approximation $\sigma_{\varphi}/\sigma_r = \kappa / 2\Omega$, where $\kappa$ is the epicyclic frequency and $\Oo \Omega \equiv V_* / r$. In general, for disc galaxies the ratio $\sigma_{z}/\sigma_r$ lies in the interval of $0.5-0.8$~\citep*{2003AJ....126.2707S,2011ARA&A..49..301V}. However, the detailed simulations of individual early-type galaxies with the dynamically hot discs, which are similar to our galaxy sample provide larger values up to $\sigma_z/\sigma_r \approx 1$~\citep{2012AstBu..67..362Z}. Dynamical models  of the individual galaxies demonstrate the variation of the  $\sigma_z/\sigma_r$ ratio along the galactic radius and this relation depends on the parameters of the galaxy~\citep{2010AN....331..731K,2012AstBu..67..362Z}. However, here in the simulations we neglect this effect. In this work we have used the ratio $0.9$ for both galaxies. Fortunately, this  choice does not critically affect our results. As a demonstration, we show in Figure~\ref{fig::circular_vels}  the circular velocities reconstructed for both galaxies
with   different values of $\sigma_z/\sigma_r$. It is clearly seen  that the uncertainties are within the $5-10\%$ scatter from the mean value.
Previous assumptions give us a possibility to reconstruct the circular velocity of stars, which corresponds to the observational galactic rotation velocity $V_c^{obs}$.

Circular velocity of the central disc or rotation velocity of the gaseous polar component unambiguously determines the total gravitational potential of the system:
\begin{equation}\label{eq::Vc}
\Oo - V_c(r_1)^2 = - r_1 \frac{\partial \Psi_{tot}}{\partial r_1} \,,
\end{equation}
\begin{equation}\label{eq::Vg}
- V_g(r_2)^2 = - r_2 \frac{\partial \Psi_{tot}}{\partial r_2} \,,
\end{equation}
where $r_1 = \sqrt{x^2+y^2}$  and $r_2 = \sqrt{x^2+(z \sin(\delta))^2}$.
The total gravitational potential is the superposition of the stellar disc, bulge, polar ring and dark matter halo potentials:
\begin{equation}
\Oo \Psi_{tot} = \Psi_{disc} + \Psi_{bulge} + \Psi_{ring} + \Psi_{halo}\,.
\end{equation}
We have used the exponential disc $\Psi_{disc}$ and King's bulge approximation $\Psi_{bulge}$ with the parameters given in table~\ref{Tabl2}. The following model was used for the polar ring contribution to the total gravity: 
\begin{equation}
\Oo \Psi_{ring}(r) = - \frac{G M(<r)}{r}\,.
\end{equation}
It is believed that the mass of the polar ring should be rather small and it is not sufficient to greatly affect the total gravity. However, for the SPRC~7 the estimated mass is $10^8$~\Msun, which gives the circular velocity of up to $50$~\kmps. In contrast, the gaseous mass of NGC~4262 is  four times smaller and its own circular velocity does not exceed $25$~\kmps.

To reproduce the observed rotation velocity, we need to constrain the parameters of the dark matter halo potential.
Two profiles of matter distribution in the halo were used for this  purpose:  NFW~model~\citep*{1997ApJ...490..493N} and isothermal halo model~\citep{1995ApJ...447L..25B}. As  it was mentioned above, copious previous research  has revealed that the halo is not spherical. So, the most general case is that the halo is an ellipsoid, and its scale length also depends on the coordinates.  Following~\citet*{2007MNRAS.377...50H} we modify the standard potential profiles  this way:
\begin{equation}
\Psi_{halo}(x,y,z) = \Psi_{halo}(\xi)\,,
\end{equation}
where
\begin{equation}
\xi = r \sqrt{\left[ \frac{x}{a_h(r)} \right]^2 + \left[ \frac{y}{b_h(r)} \right]^2 + \left[ \frac{z}{c_h(r)} \right]^2}\,.
\end{equation}
with normalization $r^3 = a_h(r)b_h(r)c_h(r)$ and $a_h(r) = r (b/a)^{-1/3} (c/a)^{-1/3}$. The kinematics of polar ring galaxies allow to directly constrain the halo parameters for two  axes: in the disc plane $a_h$ and in the perpendicular direction $c_h$, which is  equivalent to the pair of parameters  $a_h$ and $c/a$  where $a_h$ does not depend on radius $r$. Below we assume that the third halo's scale length $b_h$ is equal to $a_h$.

In the next paragraph, we describe the models of the halo  potential and the fitting methods for the kinematics of PRGs.

\begin{figure*}
\includegraphics[width=0.5\hsize]{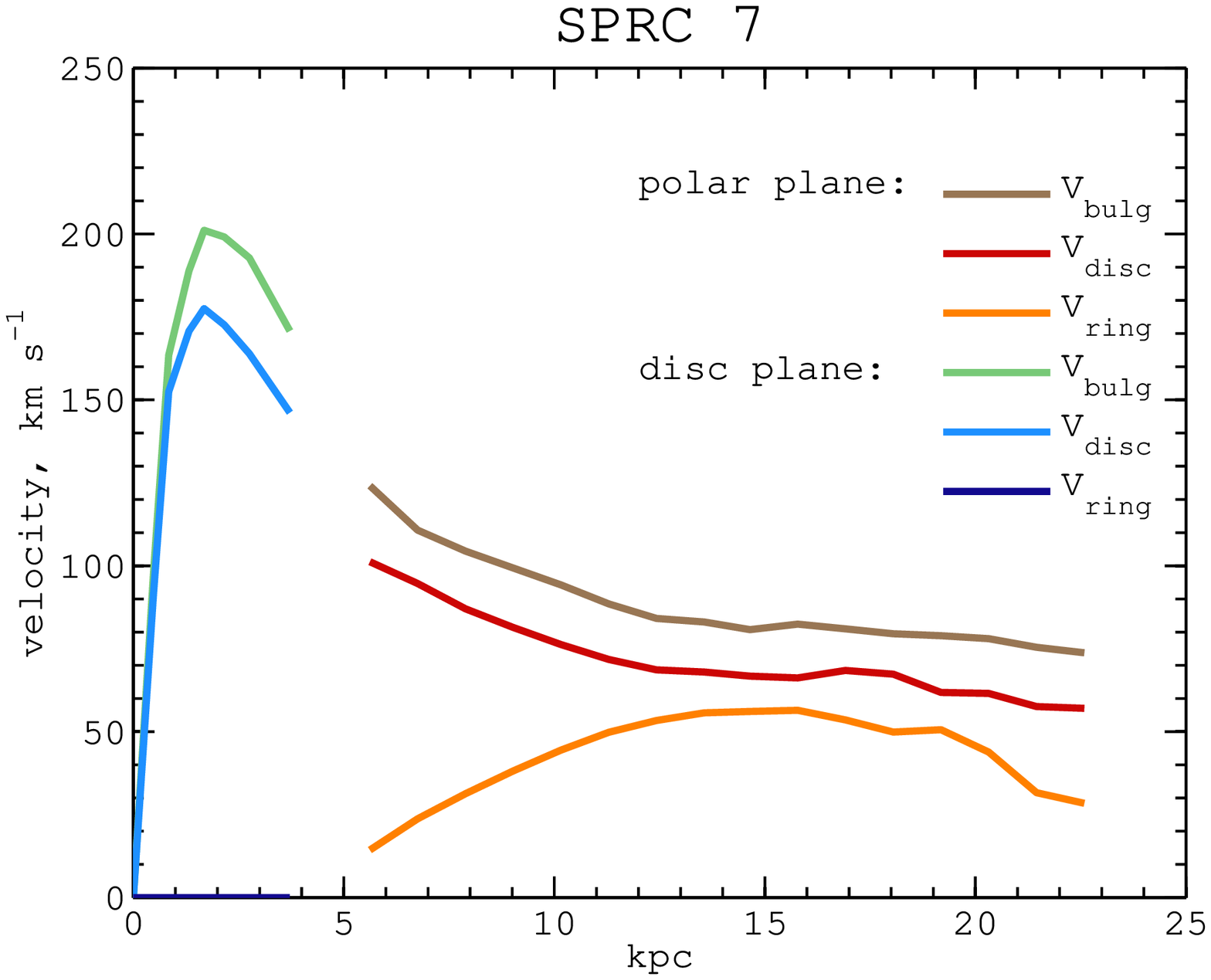}\includegraphics[width=0.5\hsize]{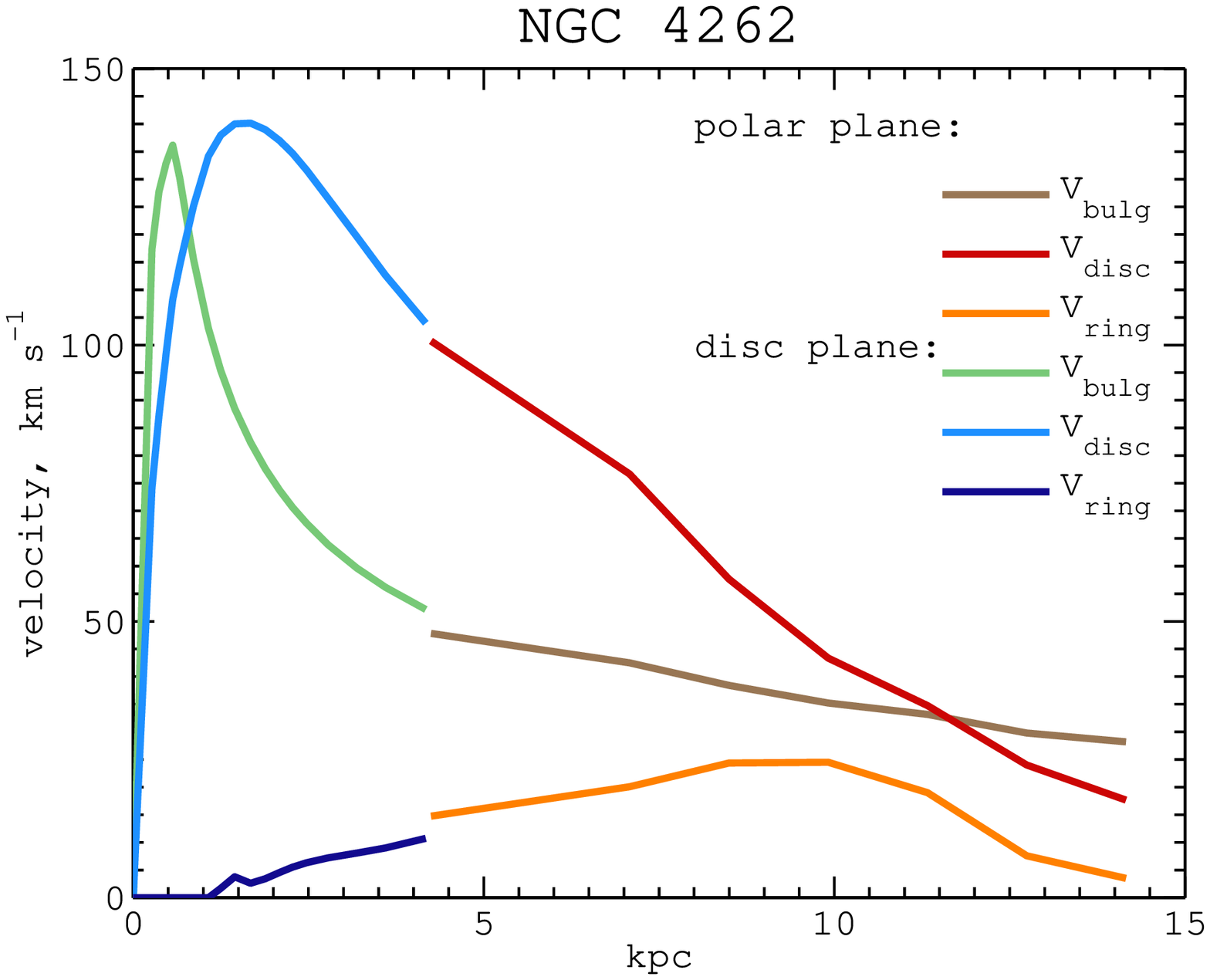}
\caption{The components of circular velocities for the bulge, the disc and the ring  in SPRC~7 (left) and NGC~4262 (right).}\label{fig::circular_decomp}
\end{figure*}

\subsection{Fitting procedure}

We can indicate three types of haloes from the previous part:
\begin{itemize}
\item Spherical halo, if $a_h = c_h$. There are two parameters: the mass of the halo $M_h$ and its scale length.
\item Oblate halo, if $a_h \neq c_h$ and we have three free parameters: $M_h$, $a_h$ and $c/a$.
\item Variable shape halo, $c/a$ is a function of radius.
\end{itemize}
The last type  requires a model of the halo axis variation. We use the following  approximation obtained for the sample of simulated haloes within the $\Lambda$CDM framework~\citep{2007MNRAS.377...50H}:
\begin{equation}
\Oo c/a = \exp{ \left[ \alpha \tanh{\left(\gamma \log{\left(r/r_{\alpha}\right)}\right)}\right]}\,,
\label{eq::c_to_a_by_Hayashi}
\end{equation}
where the  $\alpha$, $\gamma$ and $r_{\alpha}$ parameters determine the character of the halo shape.  These parameters are quite different for the potential and density distributions of the haloes, but the type of the shape variation is similar. It should be noted  that the formula describes well both the density and potential distribution of DM in the halo. Formally, such a model gives  five free parameters: $M_h$, $a_h$, $\alpha$ parametrizes the central value of the halo axis ratio and  it is given by $10^{-2\alpha}$, $r_{\alpha}$ is the characteristic radius at which the axial ratio increases significantly from its central value and $\gamma$ regulates the sharpness of the halo shape.

We examine  the different types of dark matter potential which can reproduce both the observed kinematics of the gasous polar ring and the stellar kinematics in the central galaxy simultaneously. Thus, our aim is to find the halo  parameters   producing the minimal differences:
\begin{equation}\label{eq::chi_s}
\Oo (V_c - V_c^{obs})^2 / (\delta V_c^{obs})^2 = \chi^2_s\,,
\end{equation}
\begin{equation}
\Oo (V_g - V_{gas})^2 / (\delta V_{gas})^2 = \chi^2_g\,,
\end{equation}\label{eq::chi_g}
where $V_c$ and $V_g$  are given by~(\ref{eq::Vc}) and~(\ref{eq::Vg})  correspondingly, and $\delta V_c^{obs}$, $\delta V_{gas}$ are the observational uncertainties  of the velocities. However, a more flexible approach is to minimize
the normalized superposition of $\chi^2_s$ and $\chi^2_g$:
\begin{equation}\label{eq::chi_total}
\Oo \chi^2 = \chi_s^2 / n_s + \chi_g^2 / n_g \,,
\end{equation}
where $n_s$ and $n_g$ are the  numbers of observational points on the rotational curves.

In order to fix some systematic features of the kinematics of the polar gaseous components (an elongated beam for NGC~4262, a possible spiral structure in SPRC~7) we do not fit the whole velocity fields. To convert the data from the simulated gas velocity field to the one-dimensional rotation curve, we have used a simple  assumption of quasi-circular motion in  the ring~\citep[see][and references therein]{2000AstL...26..565M}.  In the polar coordinates in the sky plane there is a relation between the line-of-sight velocity field $V(r,PA)$ and the de-projected rotation curve of gas:
\begin{equation}\label{eq::v_rot}
V(r_2,PA) = V_g(r_2) \frac{\cos(PA-PA_0)}{(1+\sin^2(PA-PA_0)\tan^2 i_1)^{1/2}}\,,
\end{equation}
where $r$ is the distance from the rotation center, $PA$ is the position angle in the sky plane and $PA_0$ is the line-of-nodes position angle. The rotation curve was calculated through the minimization of the difference between the left and right sides of the equation~(\ref{eq::v_rot}).

To take into account the azimuthal inhomogeneity of the velocity field caused by the perturbation from the central galaxy and from the non-spherical DM halo, we have simulated the  velocity field in ballistic assumption. After that we obtain $V_g$ using equation~(\ref{eq::v_rot}). Thus, we have the gaseous rotation curves from the observations, and  the simulated velocity fields as well.

The fitting procedure is accomplished using a combination of  simulated annealing and  downhill simplex methods. On every single step of the fitting we vary the parameters of the DM halo and hence circular velocities change too.

\section{Results}

\begin{figure*}
\includegraphics[width=0.45\hsize]{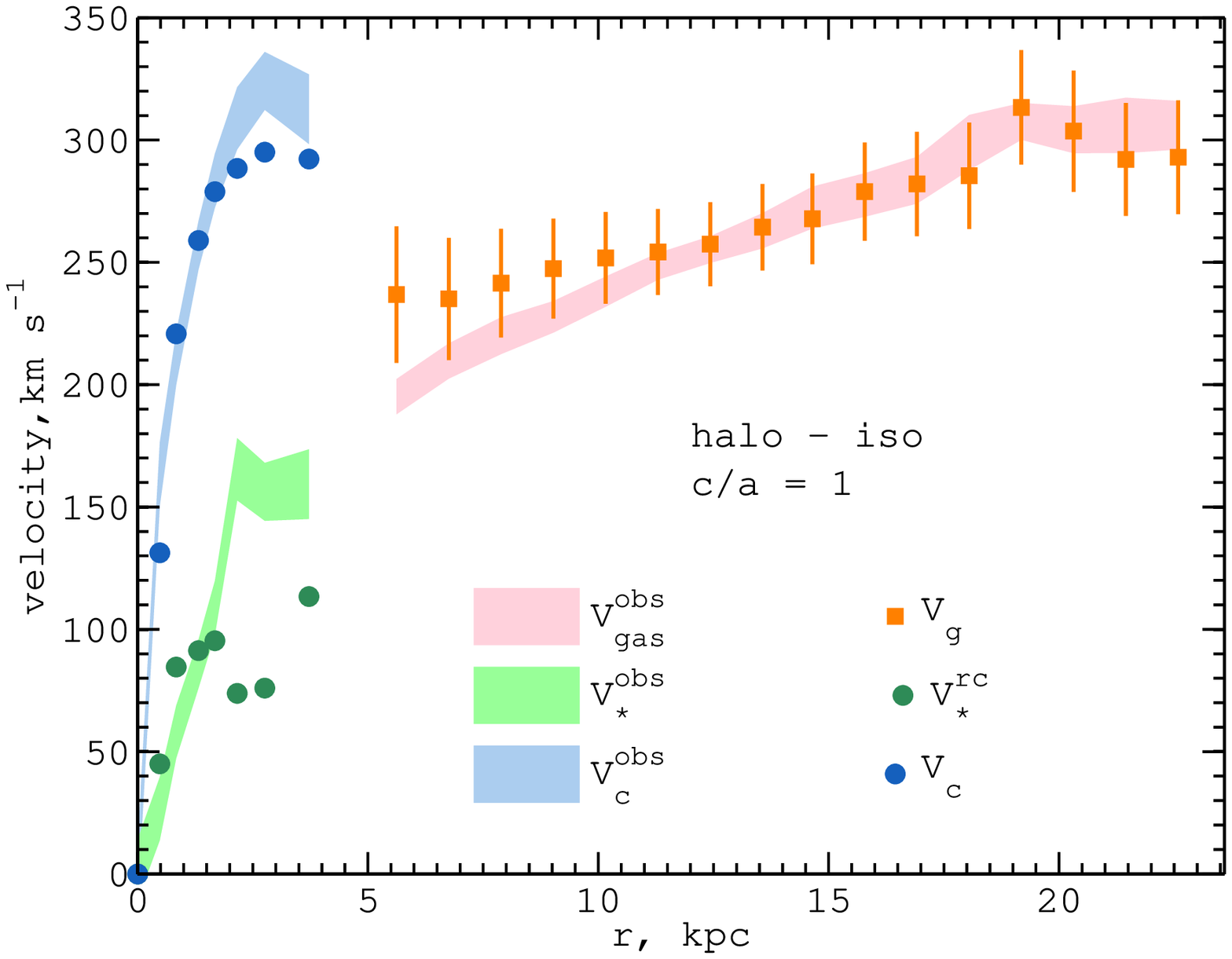}
\includegraphics[width=0.45\hsize]{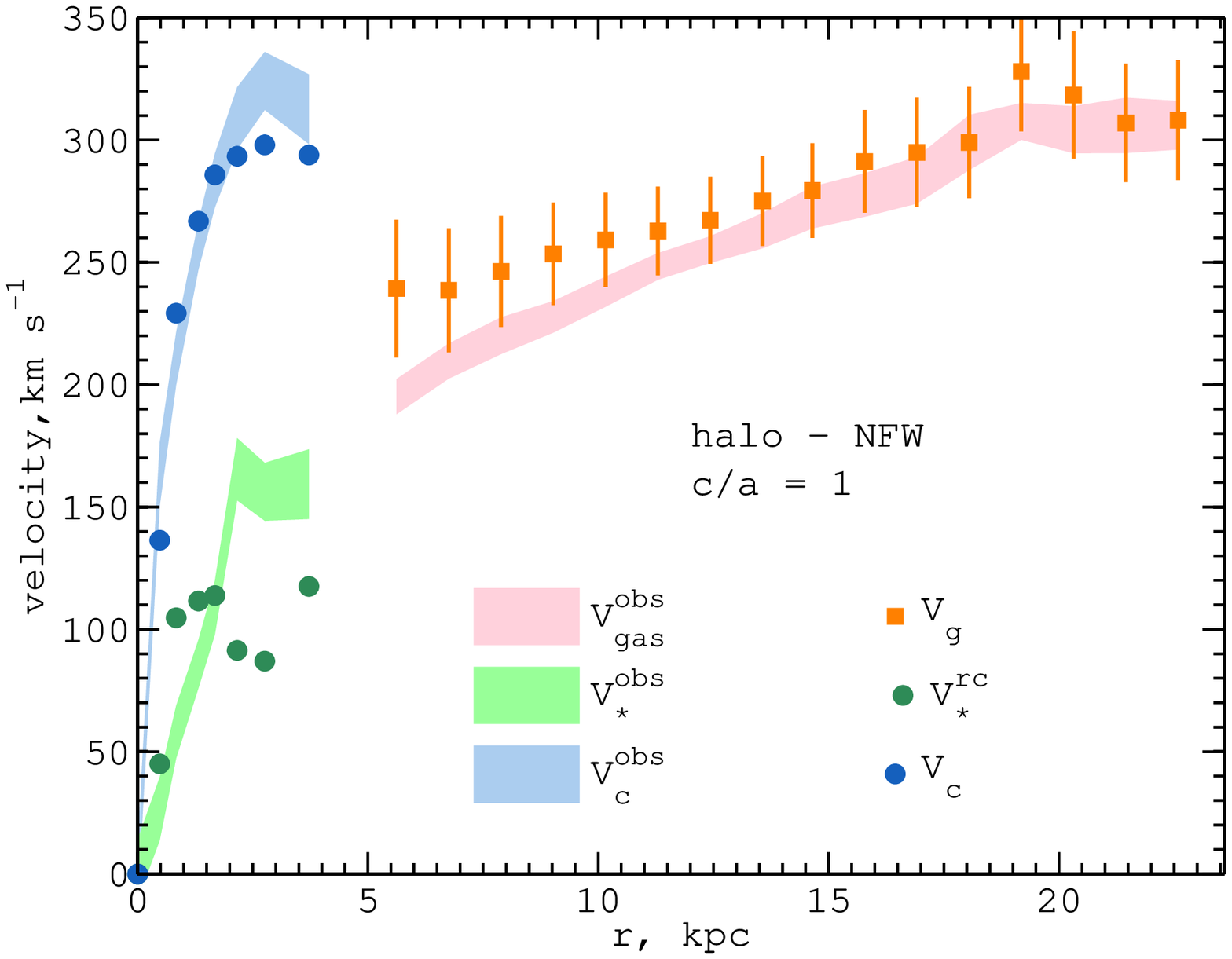}

\includegraphics[width=0.45\hsize]{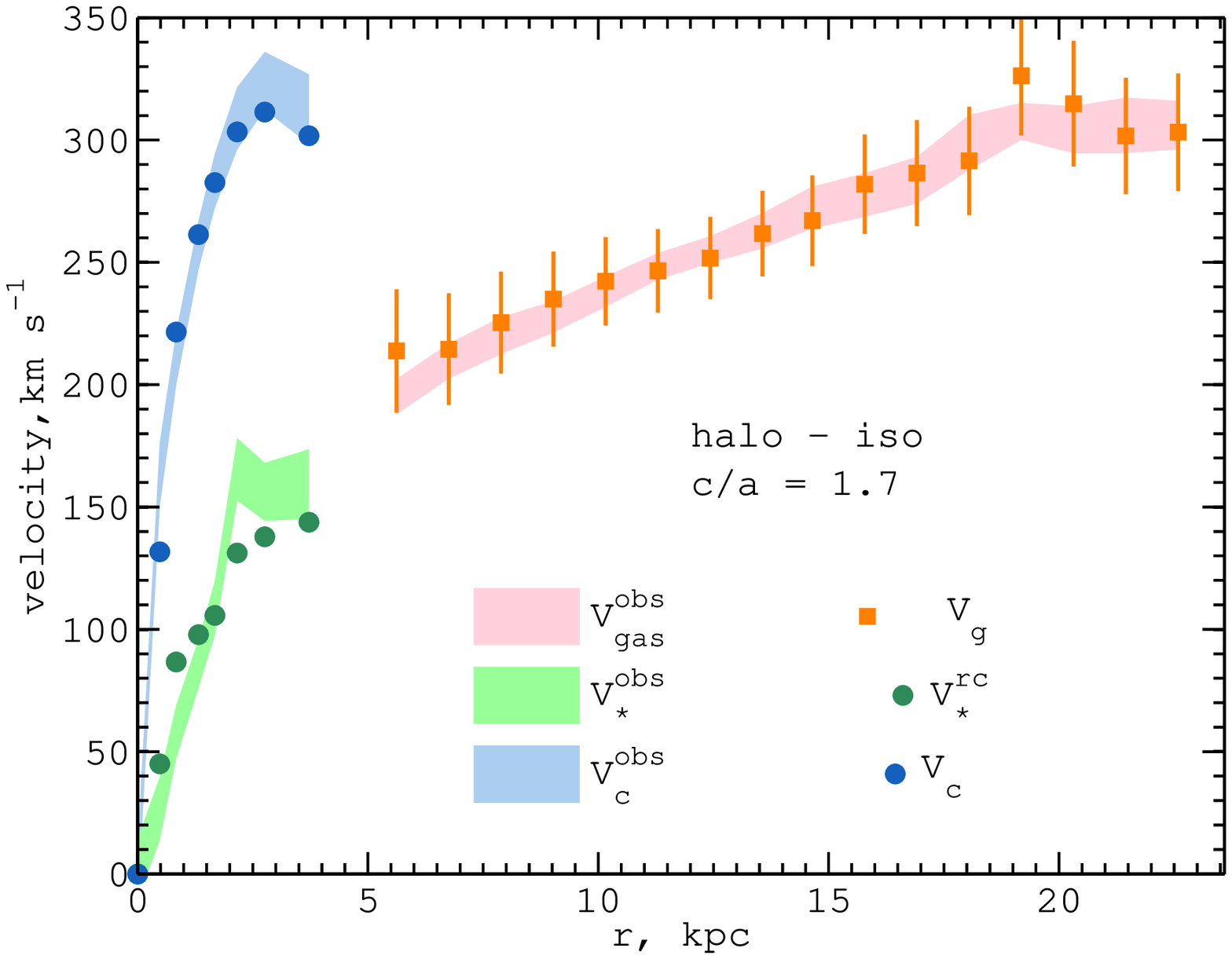}
\includegraphics[width=0.45\hsize]{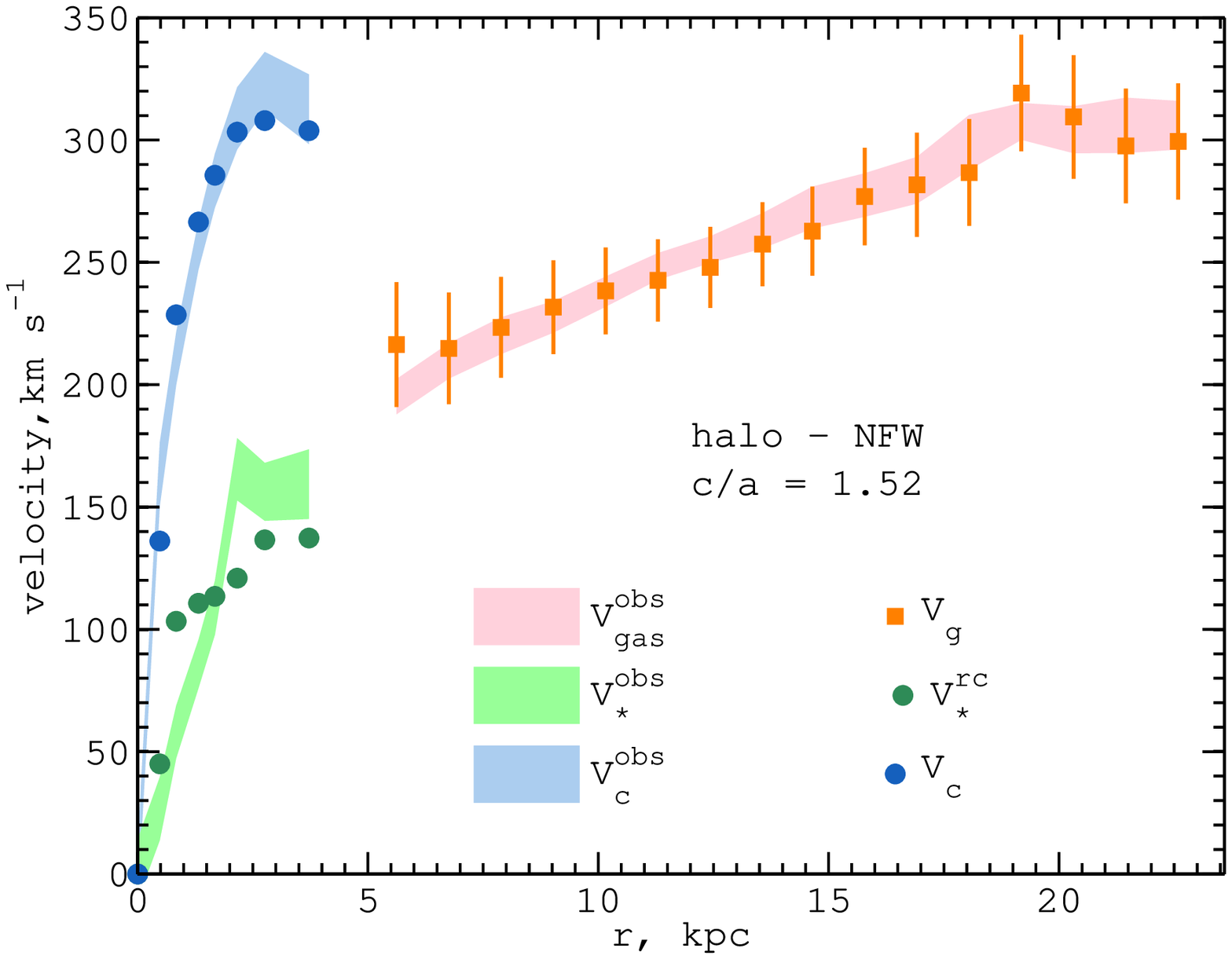}
\caption{Left column: the best-fit models of rotation velocity for both components of SPRC~7 for the isothermal halo profile.  Right column: the same, but for the NFW model. The first row corresponds to the spherical model ($c/a=1$). The second row is for the model with the constant  $c/a \neq 1$ ratio.}\label{fig::best_fit_sprc7}
\end{figure*}

\subsection{SPRC-7}

The simulations of the kinematics of this galaxy allow us to conclude that the models of spherical halo can not accurately reproduce   the rotation of both components of SPRC~7. It is clearly seen  that  these best-fit models produce a smaller circular velocity of CG and faster rotation of the polar component than those expected from the observations~(see Figure~\ref{fig::best_fit_sprc7}). This discrepancy is especially well seen  in a comparison of rotation curves of CG and models. The spherical halo model does not reproduce the flat part of the rotation curve at $r = 2 - 4$~kpc, where  the difference is about $50-70$~\kmps. This result is not surprising, because it agrees with numerous previous investigations of PRGs. We found that best-fit model of the potential for SPRC~7 is the halo flattened  towards the polar plane with the axis ratio $c/a = 1.7 \pm 0.2$ for the isothermal halo model and $c/a = 1.5 \pm 0.2$ for NFW. Figure~\ref{fig::best_fit_sprc7}   shows both of these models. The masses of haloes are also quite close to each other: $1.5 \pm 0.2 \cdot 10^{11}$~\Msun, for the isothermal halo and $1.9 \pm 0.1 \cdot 10^{11}$~\Msun\, for the NFW halo profile. Basically, the deviation of the model rotation curve from the observations is not high for all types of models (flattened or spherical halo) shown here. The general difference is in the central part of the polar ring, where the residual velocity is about $40-50$~\kmps. However, even such values are within the error bars. The distinctive feature of various solutions is in the kinematics of CG. Large values of circular velocity~(e.g. Figure~\ref{fig::circular_vels}) up to $300$~\kmps require a relatively short halo axis in the CG plane. At the same time, a slow growth of the rotation curve with radius  is provided by the large halo axis in the polar plane. The combination of both of these  aspects  gives us a flattened halo  towards the polar plane.

\subsection{NGC 4262}

\begin{figure*}
\includegraphics[width=0.45\hsize]{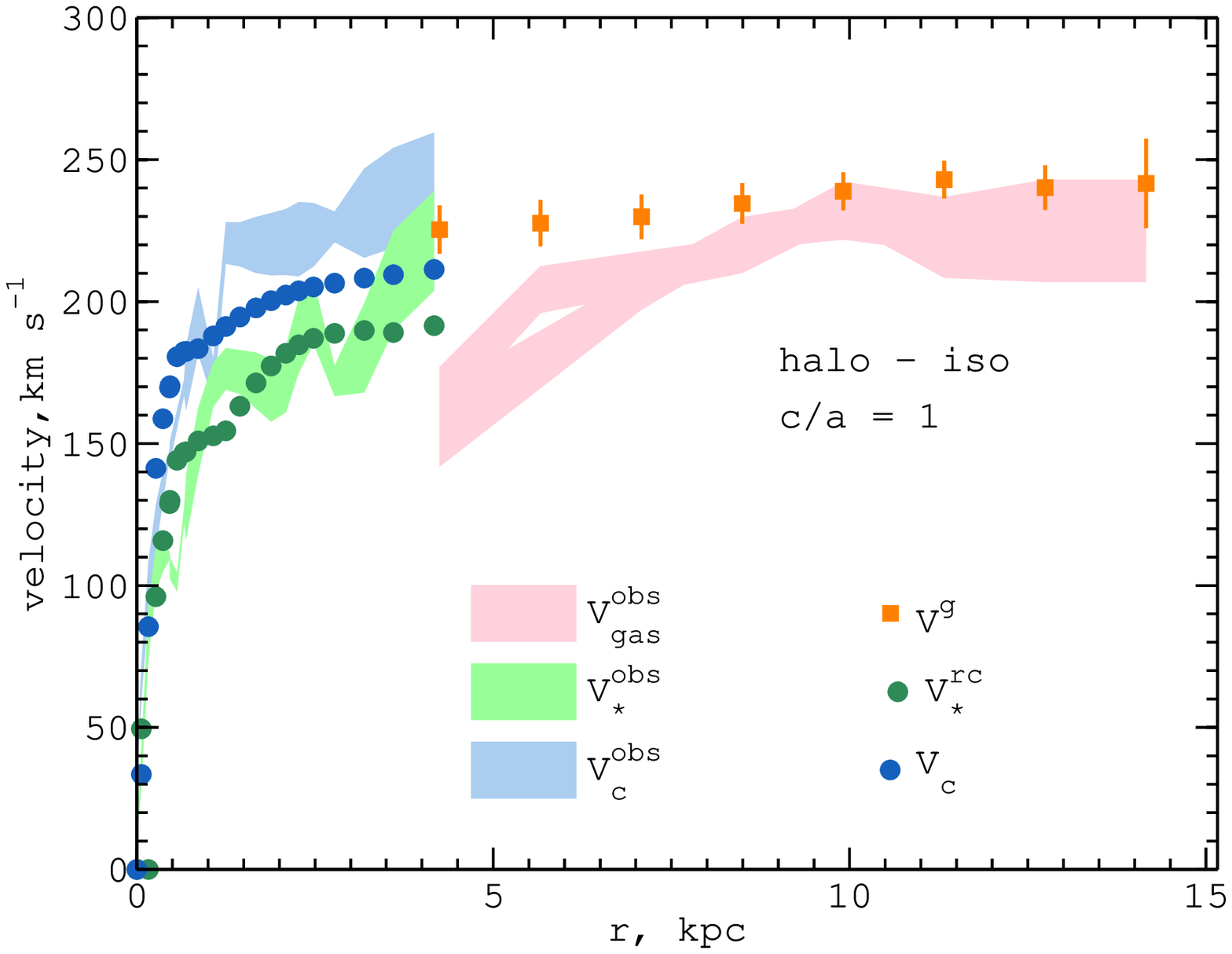}
\includegraphics[width=0.45\hsize]{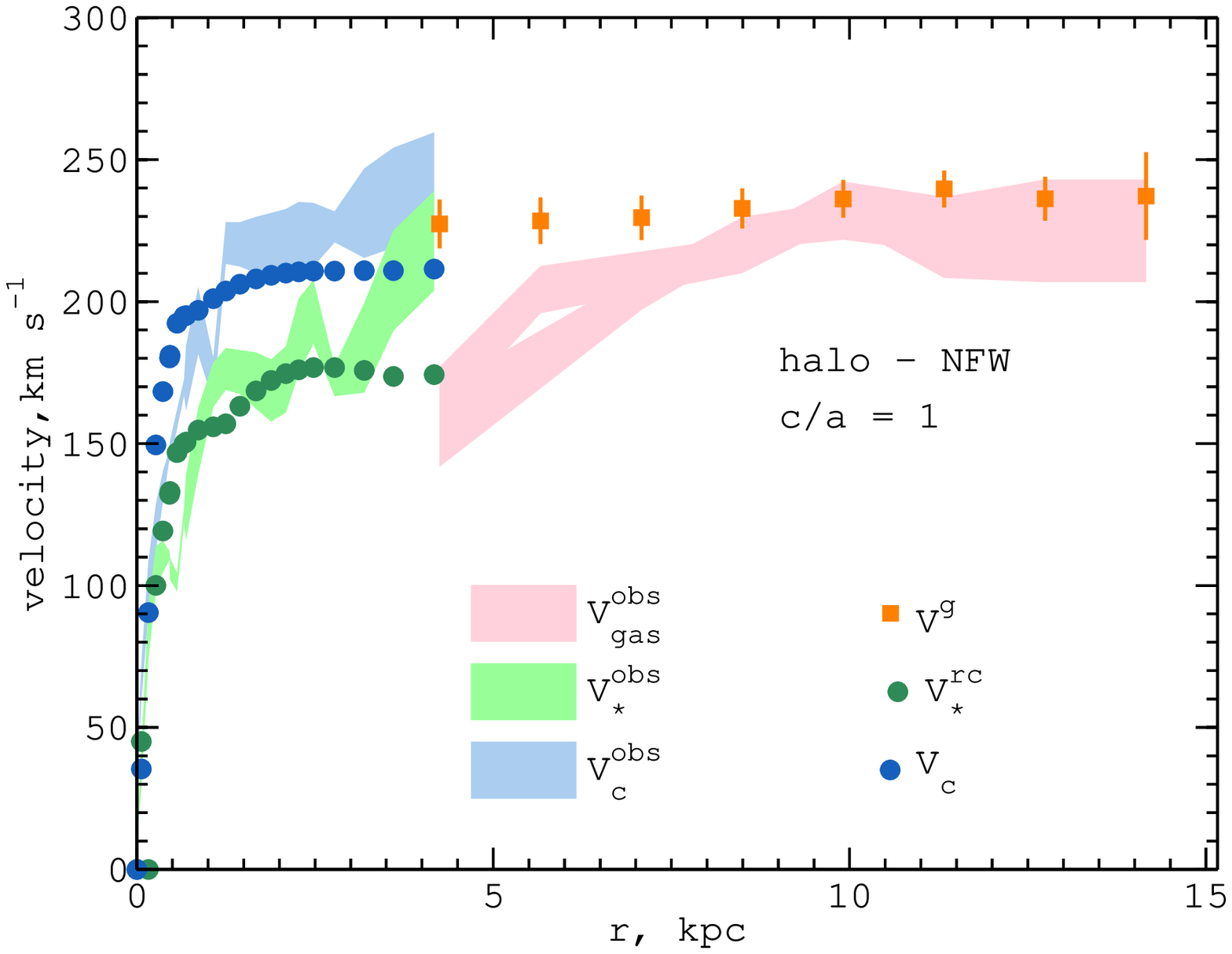}

\includegraphics[width=0.45\hsize]{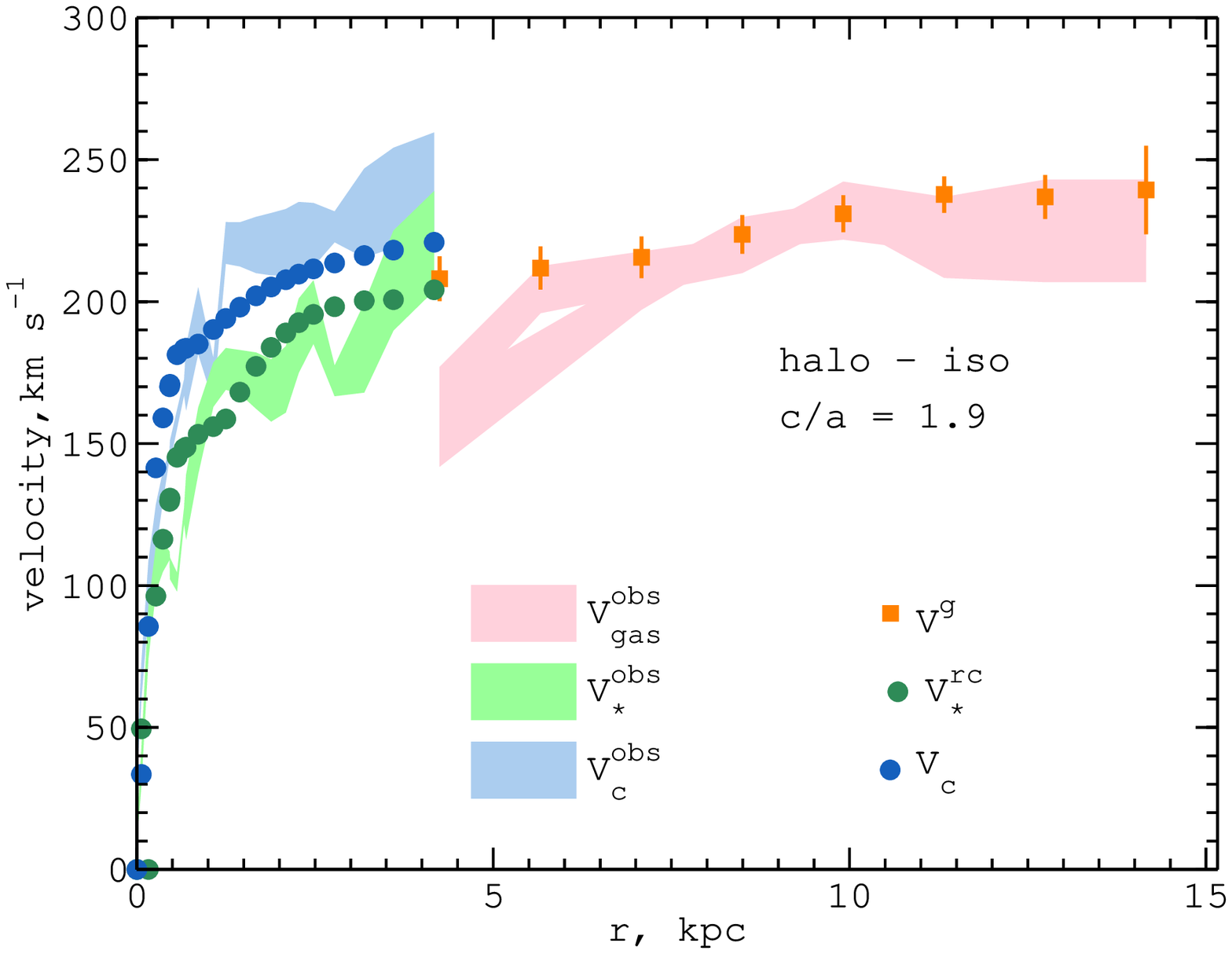}
\includegraphics[width=0.45\hsize]{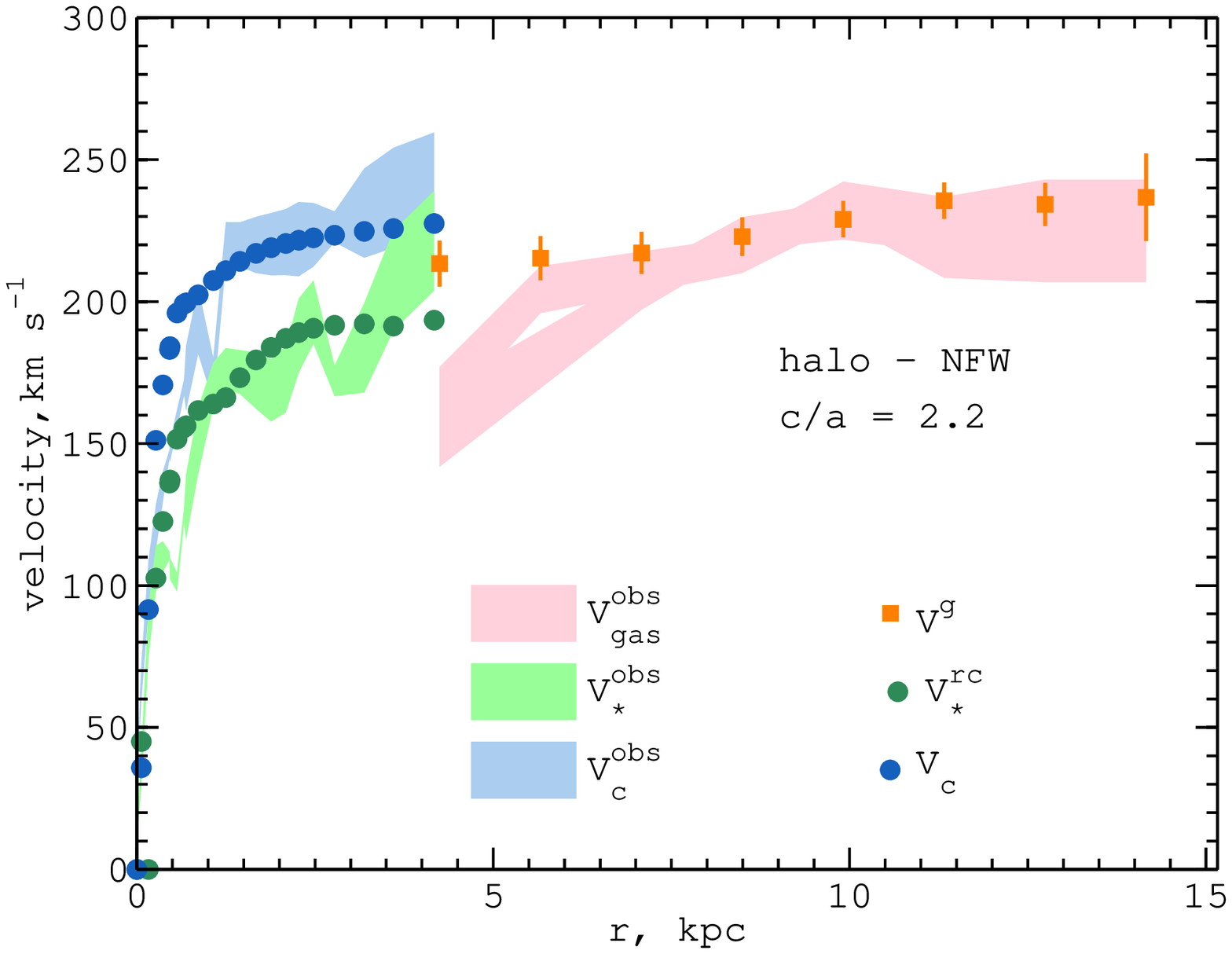}

\includegraphics[width=0.45\hsize]{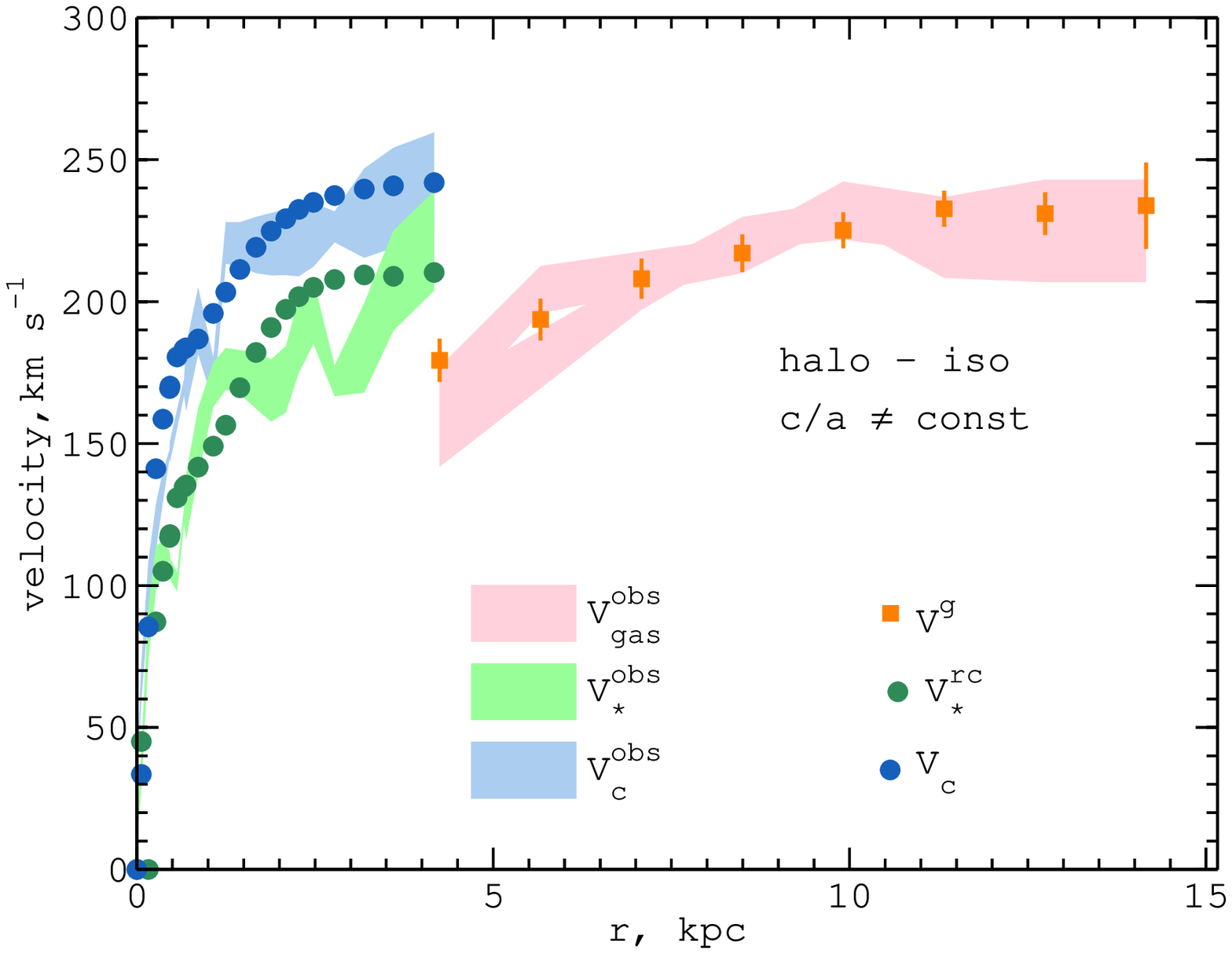}
\includegraphics[width=0.45\hsize]{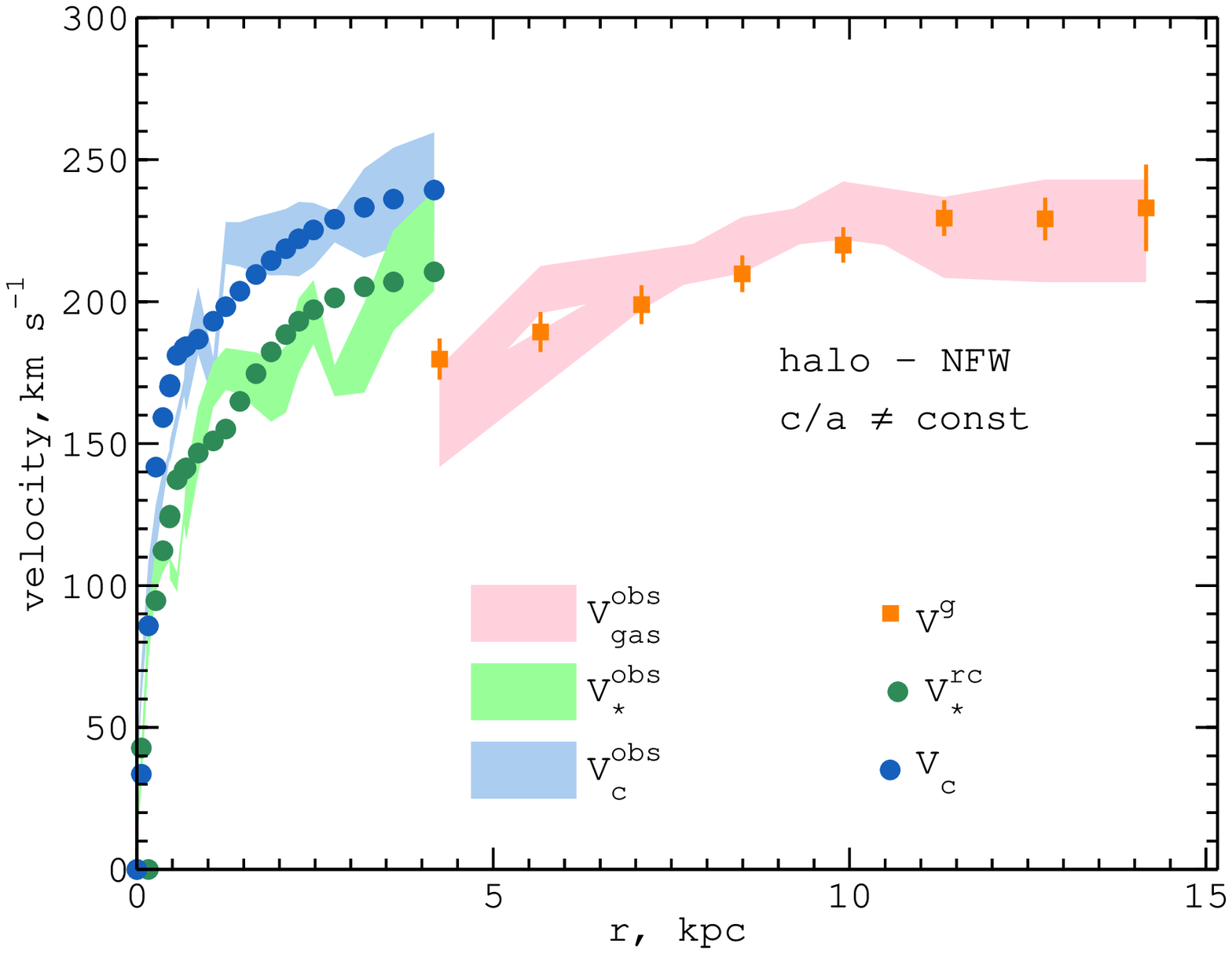}
\caption{The same as in fig~\ref{fig::best_fit_sprc7}  but for NGC~4262. The  third row is for the models with a  variable halo axis ratio. }\label{fig::best_fit_4262}
\end{figure*}

The variation of the halo mass $M_h$, halo scale length $a$ and halo axis ratio $c/a$ didn't bring a sufficient result for the reconstructed disc and ring rotation. It is well seen in the Figure~\ref{fig::best_fit_4262} that the model curves  have a bad agreement with the observations, especially for the rotation curve of the polar ring for the models with $c/a=1$ and $c/a=const$. The reason is as follows. It is obvious that circular velocity for  the central galaxy is  significantly higher than the measured rotation curve because of the asymmetric drift. The fact that the polar ring rotates in another plane  than the host galaxy, the inner part should rotate with approximately the same velocity. For instance, the maximum value of circular velocity for NGC~4262 is 240~\kmps~(Figure~\ref{fig::circular_vels}), but the inner edge of  the ring rotates at 160~\kmps~(figure~\ref{fig_obs_RC}). Moreover, in this region the ring rotates slower than the central galaxy in contrast to the SPRC~7. Here we can summarize that a simple oblate/prolate or spherical DM halo shape does not explain the observations~(see Figure~\ref{fig::best_fit_4262}).

To achieve a good agreement between the observations and model of PRG rotation, we propose a more complicated shape of DM halo following the cosmological numerical simulations~\citep{2007MNRAS.377...50H}. We study the possibility of the variable DM halo shape in the form~(\ref{eq::c_to_a_by_Hayashi}) described in the previous Section. It should be noted that following the approximation, the simulated haloes are usually prolate. However,  they become less axisymmetric and more spherical in the outer regions. It is not clear how to  translate this result directly to our units, because~\citet{2007MNRAS.377...50H} did not consider the baryon fraction and there is no opportunity to estimate the arrangement of the galactic disc in their model haloes. Furthermore, the condensation of baryons should lead to rounder halos~\citep{2008ApJ...681.1076D}. Nevertheless, we assume that  a simple approximation of the halo shape is a common case.

\begin{figure}
\includegraphics[width=1\hsize]{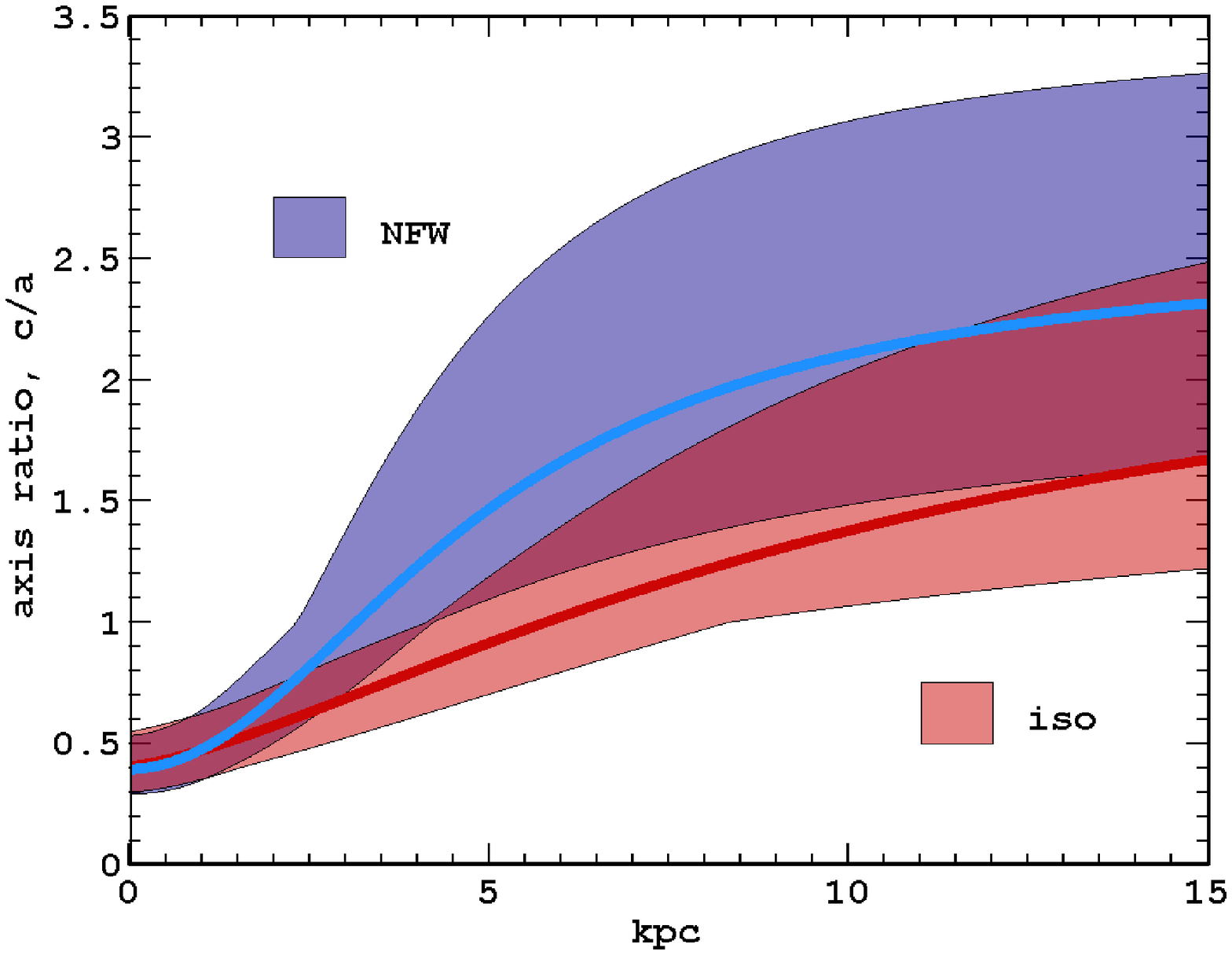}
\caption{The axial ratios of DM halo of NGC~4262 as a function of radius in the best-fit models with radially varying shapes. The solid lines are the best-fit models of the halo axis ratio according to the parameters from the Table~\ref{Tabl3}. The color regions mark the $\pm3\sigma$ range of the $c/a$ ratio for the NWF halo model (blue) and the isothermal halo (red).}\label{fig::c_to_a}
\end{figure}

We apply the fitting procedure for the kinematics of NGC~4262 accounting for the assumption of the variable shape of DM halo. This provides the 5-dimensional phase space of the parameters: $M_h$, $a_h$, $\alpha$, $\gamma$ and $r_{\alpha}$. The best-fit rotation for both components of the galaxy is shown in Figure~\ref{fig::best_fit_4262}. We have found that the DM halo is oblate within the optical radius of the central galaxy and at the same time the halo is strongly flattened  towards the perpendicular (or polar) plane far beyond this radius. The corresponding halo axis  ratio is shown in Figure~\ref{fig::c_to_a}. Both isothermal $(c/a)_{iso}$ and NFW $(c/a)_{NFW}$ models demonstrate   similar profiles, but $(c/a)_{NFW}$ has a much steeper slope. The central value is exactly the same and it  is equal to $0.4 \pm 0.1$. The isothermal halo becomes round at $r = 5.5-6$~kpc and NFW at $r = 3-3.5$~kpc, these values are quite close to the optical size of CG. At the large radii, the halo axis ratio  differs a bit: $(c/a)_{NFW} = 2.3$ versus $(c/a)_{iso} = 1.7$ at $r=15$~kpc. It is worth noting that  large uncertainties exist especially at large distances from the galactic center. This is explained by the weak sensitivity of the model from the gravitational potential at the outer part of galaxy. The inner axis ratio is strictly  smaller than unity, at the same time, despite the large error bars at the outer part halo axis ratio $c/a$ is strictly larger than unity. Thus, we should underline that such kind of halo shape is strongly required to reproduce the observational kinematics of NGC~4262.

The quantitative discrepancy between the NFW- and isothermal DM shapes, shown in Figure~\ref{fig::c_to_a} might be explained in the following way. We have fit the same kinematics by  both the NFW and isothermal halo models having slightly different masses~(Table~\ref{Tabl3}). It is possible only in case of various halo axis ratio. Namely, a relatively larger halo mass explains the given galaxy kinematics in case of a larger deviation of the DM from the spherical shape. In   Figure~\ref{fig::c_to_a} NFW best fit model have   larger values of $\Oo c/a$. Meanwhile, we can  see from   Table~\ref{Tabl3}   that the mass of the DM halo with the NWF profile is systematically larger than in the isothermal profile for all models. This relation is in agreement with the cosmological simulations by~\citep{2006MNRAS.367.1781A}, where the major-to-minor axis ratio (always less than 1) decreases with the growing virial mass of the DM halo. 

In this context, there  comes  a question about the possibility of separation of the NFW and  isothermal profiles for the given galaxies if there exist both constrains on mass and the halo axis ratio. However, our solutions  always overlap beyond the galactic center due to large uncertainties. Hopefully, a larger sample of DM shapes will yield more reliable conclusions.

\begin{table*}
\caption{Best-fit halo  parameters for the considered galaxies}
\begin{tabular}{lccccccc}
\hline
Name (halo) model & $M_h$(<14~kpc) &  $a_h$ &  $c/a$& $\alpha$  &$r_{\alpha}$ & $\gamma$ \\
     & 10$^{11}$~\Msun   & kpc &  &  & kpc &  \\
\hline
SPRC~7 (iso) I     & $1.81 \pm 0.52$ &  $1.82 \pm 0.5$ 	 & 1 & -- & -- &  --  \\
SPRC~7 (iso) II    & $1.56 \pm 0.23$ &  $1.45 \pm 0.41$	     & $ 1.7 \pm 0.22 $ & -- & -- &  --  \\
\hline
SPRC~7 (NFW) I 	   & $1.83 \pm 0.11$ &  $17.4 \pm 1.77$  	 & 1& -- & -- &  --  \\
SPRC~7 (NFW) II    & $1.9 \pm 0.11$ &  $14.6 \pm 1.64$	 & $ 1.52 \pm 0.16$ & -- & -- &  --  \\
\hline
\hline
NGC~4262 (iso) I   & $1.25 \pm  0.16$ &  $1.7  \pm  0.48$  & 1 & -- & -- &  -- \\
NGC~4262 (iso) II  & $1.2 \pm  0.2$ &  $2 \pm 0.3$  & $1.9 \pm 0.25$ & -- & -- &  --  \\
NGC~4262 (iso) III & $1.6 \pm  0.2$ &  $2.4 \pm  0.46$   & -  & $0.9 \pm 0.3$ & $1.33 \pm 0.2$ &  $0.68 \pm 0.09$ \\
\hline
NGC~4262 (NFW) I   & $1.5 \pm 0.15 $ &  $14.8 \pm 2.7 $ & 1 & -- & -- &  -- \\
NGC~4262 (NFW) II  & $1.8 \pm 0.3 $ &  $13 \pm 2.1$     & $2.2 \pm 0.7$ & -- & -- &  --  \\
NGC~4262 (NFW) III & $1.9 \pm 0.4 $ &  $11 \pm 2$     & -- & $ 0.93 \pm 0.3$ & $ 0.95 \pm 0.1$ &  $ 1.1\pm 0.2 $ \\
\hline

\end{tabular}\label{Tabl3}
\\
$M_h$ --- halo mass within $14$~kpcs;
 $a_h$ --- halo scale length along the X-axis in the host galaxy plane;
 $c/a$ --- halo axis ratio in polar and disc planes for the non-spherical halo model;
 $\alpha$, $r_{\alpha}$, $\gamma$ --- halo shape parameters in case of a variable axis ratio (see eq.\ref{eq::c_to_a_by_Hayashi}).
\end{table*}

\section{Discussion}

We have investigated  two polar ring galaxies in the context of  the shape of the dark matter halo around them. We found the first estimates of the halo axis  ratio for SPRC~7 and NGC~4262. An expectable result for  SPRC~7 is a  halo flattened towards the polar ring plane. NGC~4262 becomes a more interesting case: it was shown that the halo axis ratio varies with radius from being flattened towards the CG plane in the center and up to being prolate  towards the polar plane far beyond the CG. 
There are a few evidences that DM haloes have such kind  of shape  in our Galaxy. \citet{2011ApJ...732L...8B} justify the non-constant Milky Way halo axis ratios  using the flaring of the H~{\sc i} layer. Meanwhile, evidences of varying halo axis ratio were found from  the study of the Sagittarius stream. \citet{2013ApJ...773L...4V} have found that our Galaxy dark halo is oblate with $c/a = 0.9$ (within $10$~kpc) while the outer halo can be made mildly triaxial. It is possible that in general, the DM halo shape of the NGC~4262 is similar to the variable shape of the Galaxy.

Nevertheless, NGC~4262 is the first example  of DM variable shape in  a distant galaxy. 
Can it be proposed hypothesized in this context 
that an oblate central part and a prolate outer region is a common situation for the overwhelming amount of galaxies? It is not obvious now due to the poor sample of objects with  well-measured dark matter halo properties. At the same time, the quality of the observational data for distant galaxies is slightly limited. Note that NGC~4262 belongs to the Virgo cluster, and therefore, a strong interaction should have occurred during the galaxy formation history.

Both  galaxies considered have large velocity dispersion inside the optical radius. Perhaps the main reason for this feature is related to the history of formation and evolution of galaxies, which is reflected   in the dynamical overheating of the CG stellar component. It is likely that an external processes (gas accretion and/or tidal effects and/or mergers) are responsible for it.

Why  are the resulting shapes of DM haloes for PRGs in our study different? On the one hand, both central galaxies have similar sizes and rotation curves with a large velocity dispersion going up to 200-250~\kmps in the center. A  rather large amount of DM is needed to stabilize the dynamics of  host galaxies. 

\begin{figure}
\includegraphics[width=1.\hsize]{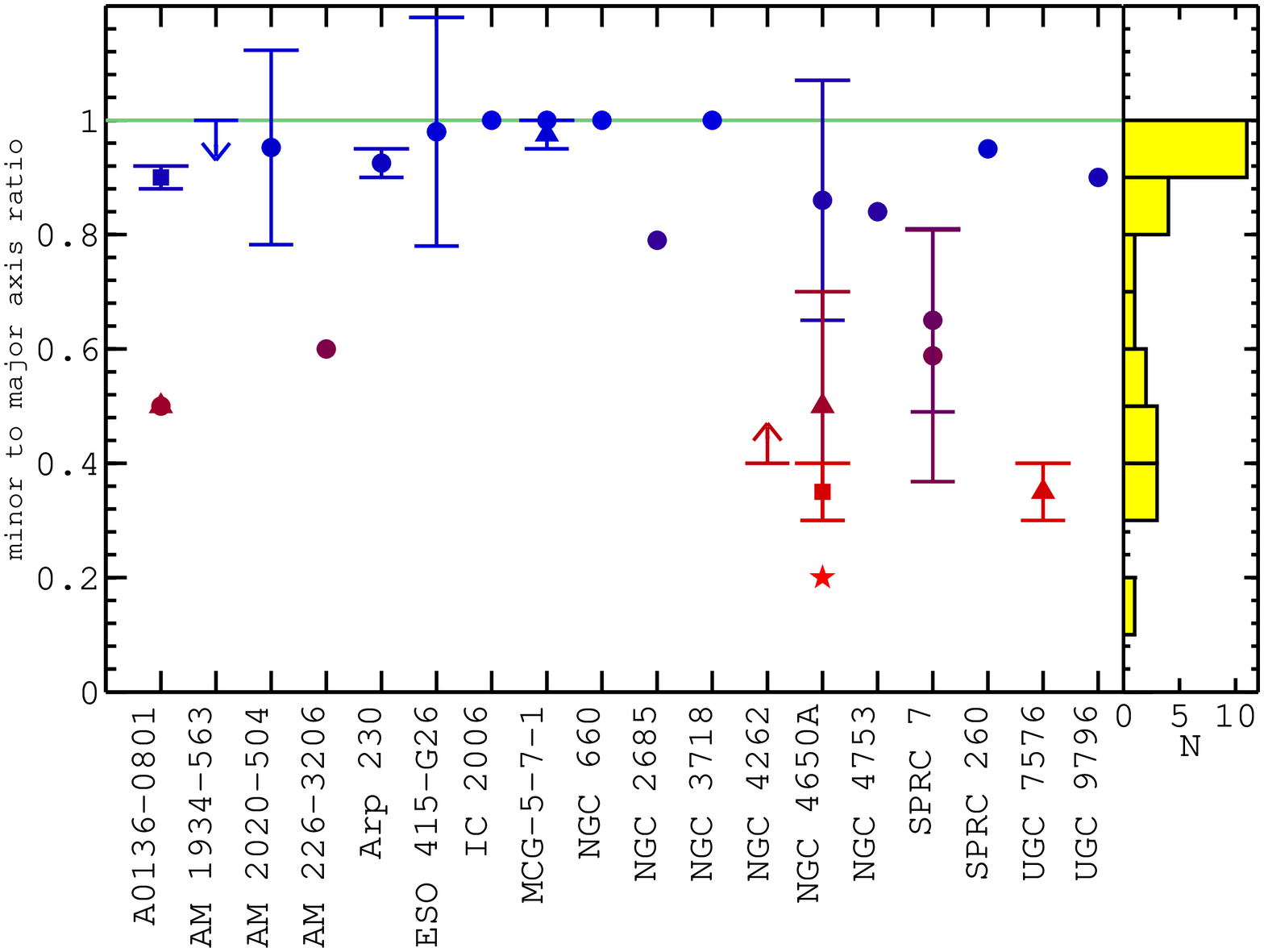}
\caption{The minor-to-major axis ratio of the DM halo obtained for different well-studied PRGs: A0136-0801~\citep{1994ApJ...436..629S,1995AIPC..336..141S},
AM~1934-563~\citep{2007MNRAS.382.1809B},
AM~2020-504~\citep{1992NYASA.675..207A},
AM~226-3206~\citep{1987ApJ...314..439W},
Arp~230~\citep{2013AJ....145...34S},
ESO~415-G26~\citep{1987ApJ...314..439W},
IC~2006~\citep*{1994ApJ...436..642F},
MCG-5-7-1~\citep{1996ASPC..106..168C, 2013AJ....145...34S},
NGC~660~\citep{1995AJ....109..942V},
NGC~2685~\citep{1993AJ....105.1378P},
NGC~3718~\citep{2009AJ....137.3976S},
NGC~4262~(this work),
NGC~4650A~\citep{1987ApJ...314..439W,1990ApJ...361..408S,1994ApJ...436..629S,1996A&A...305..763C}, NGC~4753~\citep*{1992AJ....104.1339S},
NGC~5122~\citep{1996ASPC..106..168C},
NGC~5907~\citep{2000AstL...26..277R},
SPRC~7~(this work),
SPRC~260~\citep*{2013MSAIS..25...51K},
UGC~4261~\citep*{1998ARep...42..439R},
UGC~7576~\citep{2002sgdh.conf..178S},
UGC~9796~\citep*{2006AJ....131..828C}.
}\label{fig::c_to_a_compare}
\end{figure}

However, a striking difference between these PRGs is the ring rotation character. It is clearly seen that the rotation velocity of the ring of SPRC-7 grows  from 200~\kmps up to 320~\kmps and becomes larger than the circular velocity of the S0 disc. This rotation should be provided by the DM halo with a rather large scale length in the polar plane. This expectation is confirmed by our results. For NGC~4262 the situation is not as clear because the gaseous ring component rotates at a slower rate than the circular velocity of the S0~disc. The  described pattern requires a more complicated shape of DM halo than the one obtained above.  The previous research allowed us to analyze the halo axis ratio for 18 PRGs, including the current manuscript~(see Figure~\ref{fig::c_to_a_compare}). The halo shapes were estimated using various models along with diverse observational data   (stellar and/or gaseous kinematics and/or photometrical data). Several galaxies have a few estimates for the halo shape, e.g. A0136-0801 --- 3, MCG-5-7-1 --- 2, NGC~4560A --- 4 and SPRC~7 --- 2 for different halo profiles. Generally, for the known sample of PRGs the shape of the halo is close to round $c/a \approx 1$. In the right panel of Figure~\ref{fig::c_to_a_compare}, the number of galaxies versus the minor-to-major axis ratio is shown by the yellow histogram.
It is clearly seen that there is a bimodal distribution with the mean values of $c/a \approx 1$ and $c/a \approx 0.4$. The recent statistics is certainly strongly limited. Nevertheless, the well-investigated galaxies~(A0136-0801 and NGC~4560A) demonstrate  scattered results which tend to   $c/a \approx 0.4$. A possible explanation here is that most of the  $c/a$ estimates are the spatially averaged values, which are close to the round halo. As it was mentioned above, the expected shape of the DM halo should be more complex. However, the limitation arising from model assumptions and observational difficulties, which occurs due to the peculiar character of PRGs does not allow to carefully measure the halo shape  for the majority of PRGs.

The fact that the polar component in many  PRGs contains numerous H~{\sc ii} regions indicates the ongoing large-scale star formation process. In massive rings, which are the gravitationally unstable tightly wound spiral structures can be developed~\citep{2006A&A...446..905T}. However, most of the rings are not massive enough  for the gravitational instability, which might compress the gas up to the critical values. In this case, non-axisymmetric perturbations should play a decisive role. Effective perturbation sources are a massive central galaxy and a non-axisymmetric DM halo. Some authors were  investigating the reaction of the gaseous disc to the presence of a  non-spherical DM halo~\citep{2002ApJ...574L..21B,2003ApJ...586..152M,2012ARep...56...16K}. These results should be directly projected to the polar ring galaxies.

\section{Acknowledgements}
The authors thank  the anonymous referee for helpful comments.
We would like to thank Tom Oosterloo, who has kindly provided the processed H~{\sc i} data cubes of NGC4262. We are grateful to Gyula J\'ozsa  for    providing  the TiRiFiC software and technical advise on its usage. This work was supported by the RFBR grants  no.~12-02-31452, 13-02-00416 and by the  ``Active Processes in Galactic and Extragalactic Objects'' basic research program of the Department of Physical Sciences of the RAS OFN-17. S.K., A.M., and A.S. are  also grateful for the financial support of the `Dynasty' Foundation.  The observations obtained with the 6-m telescope of the Special Astrophysical Observatory of the Russian Academy of Sciences  were carried out with the financial support of the Ministry of Education and Science of Russian Federation (contracts no. 16.518.11.7073 and 14.518.11.7070).
This research has made use of the NED database, which is operated by the Jet Propulsion Laboratory, California Institute of Technology, under the contract with the National Aeronautics and Space Administration.The funding for the SDSS has been provided by the Alfred P. Sloan Foundation, the Participating Institutions, the National Science Foundation, the United States Department of Energy, the National Aeronautics and Space Administration, the Japanese Monbukagakusho, the Max Planck Society and the Higher Education Funding Council for England.

\bibliography{mypolarbib}
\end{document}